\documentstyle[12pt,epsfig]{article}
\newcommand{\appendixA}{\setcounter{equation}{0}
\def\theequation{\rm{A}.\arabic{equation}}\section*}
\newcommand{\appendixB}{\setcounter{equation}{0}
\def\theequation{\rm{B}.\arabic{equation}}\section*}
\newcommand{\m}{\medbreak}
\newcommand{\no}{\noindent}
\newcommand{\EQ}{\begin{equation}}
\newcommand{\eq}{\end{equation}}
\newcommand{\EQA}{\begin{eqnarray}}
\newcommand{\eqa}{\end{eqnarray}}

\newcommand{\hu}{\hspace{1.cm}}

\newcommand{\ALL}{\mbox{$A_{LL}\ $}}
\newcommand{\ALLPV}{\mbox{$A_{LL}^{PV}\ $}}

\newcommand{\r}{$\rightarrow \;$}

\def\subfigureA#1{
\leavevmode
\hbox{#1}
}
\def\rp{$R_p \hspace{-1em}/\;\:$}
\def\pr#1#2#3{ Phys. Rev. {\bf{#1}} (#2) #3}
\def\prl#1#2#3{ Phys. Rev. Lett. {\bf{#1}} (#2) #3}
\def\pl#1#2#3{ Phys. Lett. {\bf{#1}} (#2) #3 }
\def\prep#1#2#3{ Phys. Rep. {\bf{#1}} (#2) #3}
\def\np#1#2#3{ Nucl. Phys. {\bf{#1}} (#2) #3}
\def\zp#1#2#3{ Z. Phys. {\bf{#1}} (#2) #3}

\begin{document}
\begin{titlepage}
\vspace{0.2in}
\vspace*{1.5cm}
\begin{center}
{\large \bf Hunting for Contact
Interactions at HERA with polarized lepton and proton beams
\\} 
\vspace*{0.8cm}
{\bf J.-M. Virey}{$^1$}  \\ \vspace*{1cm}
Centre de Physique Th\'eorique$^{\ast}$, C.N.R.S. - Luminy,
Case 907\\
F-13288 Marseille Cedex 9, France\\ \vspace*{0.2cm}
and \\ \vspace*{0.2cm}
Universit\'e de Provence, Marseille, France\\
\vspace*{1.8cm}
{\bf Abstract \\}
\end{center}
We explore the discovery and analysis potentials of the
HERA collider, with and without
polarized beams, in search for electron-quark 
contact interactions in the neutral current channel. 
We find that the sensitivity to contact interactions when both beams
are polarized is similar to the unpolarized case, and 
is better than in the case  where one has only lepton polarization.
We emphasize that the measurement of 
spin asymmetries in such a polarized context could give some crucial informations
on the chiral structure of these postulated new interactions.
The experimental conditions are carefully taken into account.
\vspace{0.5cm}

\vfill
\begin{flushleft}
\m\no
Number of figures : 3\\

\m\no
May 1998\\
CPT-98/P.3541\\
\m\no
anonymous ftp or gopher : cpt.univ-mrs.fr

------------------------------------\\
$^{\ast}$Unit\'e Propre de Recherche 7061

{$^1$} 
E-mail : Virey@cpt.univ-mrs.fr
\end{flushleft}
\end{titlepage}

\section{Introduction}
  
\indent 
It is commonly assumed that the Standard Model (SM) is only a low energy effective
theory of a more fundamental and complete theory, which
will be able to resolve naturally several problems inherent to the SM.
For instance, as a non-exhaustive list of these difficulties let
us mention : {\it i)} the scalar structure of the SM : nature of
the Higgs boson and origin of the electroweak symmetry breaking;
{\it ii)} the huge number of SM free parameters to be fixed by
experiments; {\it iii)} the origin of parity violation for weak interactions;
{\it iv)} the origin of the SM three generations, {\it i.e.} of the apparent
replication of quarks and leptons.

Besides, the ultimate unification of all particles and of all 
interactions is still an essential aim of particle physicists. Then, it is
natural to have several models or theories which go beyond the SM, 
in order to satisfy this unification goal and to resolve simultaneously
some problems of the SM like those mentioned above.\\
The models derived from String theories or/and the Grand Unified Theories
lead, in general, to the unification of the interactions. 
Elementary particles unification leads us to suppose that there is a deep unity
between quarks and leptons and also between fermions and bosons, leading
to the ideas of substructure or/and of supersymmetry.\\

A common prediction to all these models is the existence of new exotic
particles, and, in particular, of new bosons ({\it i.e.} $Z'$, vector
and scalar leptoquarks, superpartners ...). In the case of compositeness
no particular models have emerged so far and the substructure energy scale
$\Lambda$ is, in general, not directly accessible at 
present or planned high energy collider experiments. However,
phenomenologically at some energy scale, much lower than the characteristic
scale of the underlying interaction, the presence of
common subconstituents follows by assuming a Contact Interaction (CI) \cite{ELP83,EHLQ84}, independently of the details of the model.
Moreover, it appears that the exchange of a new boson in a SM process
can also be represented by a CI, if the boson mass $M$ is far
above the center of mass energy of the elementary process
({\it i.e.} $M \gg \sqrt{\hat{s}}$). To the best of our knowledge, for
$\; lq \rightarrow lq\; $ scattering, this was first noticed in
\cite{HMShera91,Donch}, and taken into account recently in \cite{Bargerct2,pcjmv}.

So, the study of CI is a powerful way to probe new physics phenomena
in general.
The connexion between CI and compositeness is direct.
For a new
exchanged particle like a SM-like boson ({\it i.e.} $Z'$, $W'$)
or a leptoquark-like boson (scalar or vector leptoquarks, some sfermions
with $R$-parity violating couplings), the connexion is
indirect in the sense that we do not consider the resonant effect.\\

The aim of this article is to present a purely phenomenological analysis
on the experimental signatures of CI from the measurements of cross
sections and spin asymmetries in the Neutral Current channel for
lepton-proton collisions at HERA\footnote{For an unpolarized analysis
in the Charged Current sector at HERA see \cite{Donch}}.
This work complements previous reports \cite{Ruckl84,Cash,Martyn87}, 
takes into account realistic experimental conditions and extends the 
former analysis to the two-spin case, namely when both 
lepton and proton beams are simultaneously polarized. The lepton polarization option
was considered at an early stage of the HERA project (see for example
\cite{hera87,hera91}) and it is already in use at DESY with the HERMES
experiment \cite{hermes}. The proton polarization option at HERA \cite{Felt}
has been considered only recently, and it 
has motivated all along 1997, a workshop \cite{DeRoeck} where the main results
of the present analysis have been presented \cite{jmvhera97}.\\

Consequently, we will be interested in CI which belongs to the electron-quark
sector. The conventional effective Lagrangian has then the form \cite{Ruckl84}:
\EQA\label{Leq}
{\cal L}_{eq}^{NC} = \sum_q ( \eta^q_{LL}(\bar e_L \gamma_{\mu} e_L) 
(\bar q_L \gamma^{\mu} q_L) +  \eta^q_{RR}(\bar e_R \gamma_{\mu} e_R) 
(\bar q_R \gamma^{\mu} q_R)  \nonumber 
  \\   + \eta^q_{LR}(\bar e_L
\gamma_{\mu} e_L)  (\bar q_R \gamma^{\mu} q_R) + \eta^q_{RL}(\bar e_R
\gamma_{\mu} e_R)  (\bar q_L \gamma^{\mu} q_L) )  
\eqa

\no with $\eta^q_{ij}= \epsilon {g^2}/{(\Lambda^q_{ij})^2}$ where
$q$ indicates a quark flavour and $i,j$ correspond to different
chiralities, Left (L) and Right (R). $\epsilon=\pm 1$ is a sign which 
characterizes the nature of the interferences of the contact terms
with SM amplitudes. 
The coupling constant $g$ is normalized to $g^2=4 \pi$, since such
CI have been initially invoked for compositeness studies, where the
new binding forces are assumed to be strong.

If we restrict ourselves to the light quark flavours of the first generation
({\it i.e.} $q=u,d$), we find that there are eight independent terms
appearing in eq.(\ref{Leq}). In fact, we can reduce the number of
independent coefficients by the use of symmetries. For instance, the
$SU(2)_L$ symmetry indicates the relation\footnote{Note that recently it 
has been advocated \cite{DiBart,Nelson}
that the invariance under this symmetry also imposes $\eta^u_{LL}=\eta^d_{LL}$,
but this relation is no longer valid if we consider the most general
$SU(2) \times U(1)$ invariant contact term Lagrangian, due to the presence
of an isospin triplet exchange term \cite{Cash,Bargerct2}.}
$\eta^u_{RL}=\eta^d_{RL}$ \cite{Cash,Bargerct2,EHLQ84}.
If we require the presence of some higher symmetries we will obtain similar relations
among the $\eta$ coefficients, and the interested reader can find more 
details in references \cite{Bargerct2,Nelson}. Here we will not consider any particular
symmetry in order to perform a "model independent" analysis. Nevertheless,
for simplicity, we will assume some "universality" between the $u$ and $d$
quarks contributions, namely $\eta^u_{ij}=\eta^d_{ij}=\eta_{ij}$.
However, as long as a proton beam is used at HERA, one tests essentially
the structure and interactions of the $u$-quark, and more precisely its
valence component. 

With the assumption of "$u$-$d$ universality", the four "chiralities" $LL$, $RR$, 
$LR$ and $RL$, along with the sign $\epsilon$, define {\it eight} individual
models (associated to eight distinct behaviors). The CI could correspond
to one of these models or to any combination of them.
From now, we label each individual model by its non zero $\eta^{\epsilon }_{ij}$
coefficient, associated to the "chirality" $IJ^\epsilon $, where we have
added $\epsilon$ as an index for clarity.
\\

The contact terms are constrained by several experiments involving
electron-quark interactions, like for instance unpolarized and polarized deep
inelastic scattering, Drell-Yan lepton pairs production, atomic parity violation
and hadron production in $e^+e^-$ collisions. A global study of the $eq$ CI for these
processes, with the corresponding constraints on $\Lambda$, has been performed
in \cite{Bargerct2}. In particular, the atomic parity violation experiments
on Cesium atoms give some bounds of the order of $\Lambda \sim 10\; TeV$
for the individual models \cite{Bargerct2,aldo}. However, it appears that it is
easy to find some combinations of the chiralities which evade these constraints.
The symmetries which are able to realize such cancellations do not necessary
correspond to Parity Conserving (PC) interactions \cite{Nelson}. At this point,
it is interesting to remark that in the framework of composite models, some
naturalness arguments \cite{tHooft} indicate that the composite strong
dynamics has to respect some unbroken chiral global symmetries, which leads
to Parity Violation for the new interaction \cite{EHLQ84}. Moreover,
the fact that the new interaction is Parity Violating (PV) seems to be
natural in the sense that the new energy scale $\Lambda$ is above the
characteristic energy scale $v \sim 246\; GeV$ of weak interactions,
namely in an energy domain where SM PV interactions are already 
present \cite{EHLQ84}. In addition, if we examine the fermionic couplings
of the different $Z'$ usually involved in traditional searches (see
\cite{CveticGod} for a review), we can see that they are of PV nature,
in general.

Coming back to the experimental bounds on $\Lambda$, we have noticed that
the individual models are severely constrained, which is not the case when several
terms of different chiralities are involved simultaneously. In this latter case,
the authors of ref.\cite{Bargerct2} found that the present
bounds on $\Lambda$ are of the 
order of $3-4\; TeV$. Note also that the constraints originating from atomic
parity violation experiments, could also be relaxed 
if there are compensating contributions
coming from more than one new physics source \cite{Bargerct3}.
Nevertheless,
for simplicity, in the following analysis we will consider the eight models
individually. We will comment on the effects of
some models with a more complicated chiral structure 
at the end of the last section.\\

Finally, the H1 and ZEUS collaborations at HERA have observed an excess of
events, in comparison with the SM expectations, at high $Q^2$, in the deep
inelastic positron-proton cross section $\sigma_+ \equiv d\sigma / dQ^2
(e^+p \rightarrow e^+ X)$ \cite{HERAe}. Many explanations have emerged since 
these observations (see \cite{Alt1} for a nice review). 
This excess could be interpreted as a manifestation
of new physics : leptoquarks, squarks with
$R$-parity violation (\rp ) or CI. 
In this latter case, up to some difficulties \cite{Alt1}, it corresponds
to a new interaction in the up-quark sector for 
a scale $\Lambda \sim 3-4\, TeV$ \cite{Bargerct1,DiBart,Alt1}.
Since the lepton beam is made of positrons, the
cross section $\sigma (e^+p)$ is sensitive to the chiralities $LR^\pm$ and/or $RL^\pm$
where $\pm$ corresponds to $\epsilon$. We can remark that the required $\Lambda$
for the HERA excess of events, is close to the present sensitivity 
of the other experiments
studied in \cite{Bargerct2}.
However, it
appears that with the new data \cite{HERAe2} "the evidence for the signal
remains meager" \cite{Alt1}. Then we do not insist anymore on this anomaly.
We will just present the different curves of our analysis for the ``special''
value $\Lambda =4\, TeV$.

We will present in section 2 our strategy of analysis and the relevant experimental parameters.
In section 3, we will define the most powerful observables and 
discuss the discovery reaches on
the CI scale $\Lambda$. At this stage, we will consider three different running experimental
conditions : {\it i)} unpolarized, {\it ii)} lepton polarization only
(one-spin), {\it iii)} lepton plus proton polarizations (two-spin).
In section 4 we will present a strategy which could be helpful
to strongly constrain the chiral
structure of the possible new interaction. We give our conclusions in section 5.
In appendix A, we define the double polarized cross sections, 
along with the expected behavior of the most relevant observables
involved in this analysis. 
In appendix B, we present an efficient way to take into account 
the degrees of beam polarization, involved in any polarized experiments, and
to define correctly the different spin asymmetries and their associated
statistical errors.


\section{Observables and experiments}
\subsection{Preamble}

\indent
The effective Lagrangian described by eq.(\ref{Leq}) is added 
to the SM one, and allows the calculation of the differential
inclusive cross section ${d\sigma_t}^{\lambda_e \lambda_p}/{dQ^2}$ 
for the polarized deep inelastic
scattering process :
\EQ
{\vec{e}}^{\, t} \vec{p} \rightarrow e^t X \;\; ,
\eq 

\no where $t$ is the electric charge of the colliding lepton and 
$\lambda_e$, $\lambda_p$ are the helicities of the lepton and
the proton.
We have performed the
integration over the $x$ variable in order to increase the statistics, since we
are concerned with the high $Q^2$ domain. Another reason to do that, 
comes from the fact that an effect due to a CI will present
a continuum in the $x$-distribution\footnote{For a leptoquark or a
{\rp }-squark of accessible mass at HERA, the $x$-distribution
will be crucial to observe the resonance effect, since such particles
are produced in the $s$-channel.}. The expressions for the double polarized
differential cross sections are presented in appendix A. 
\\

Now, since we have defined the basic observable 
${d\sigma_t}^{\lambda_e \lambda_p}/{dQ^2}$,
we can go into the details of our analysis. This study is divided 
into three parts, corresponding 
to three different modes of running for the HERA machine. The first one is 
operating presently, 
the other ones are in 
project : {\it i)} Unpolarized; 
{\it ii)} Lepton polarization only ($e^-$ and $e^+$) (one-spin); 
{\it iii)} Lepton and proton polarizations (two-spin).\\
\no We can define several classes of observables : the cross sections 
(called $\sigma$) and the spin asymmetries. The ``pure'' spin asymmetries,
defined for each type of lepton separately, are noted $A$.
If both electron and positron beams are available, we can
define a third class of observables : the charge asymmetries (called $B$). We put in this
class the "mixed" asymmetries which are simultaneously spin and charge asymmetries.
\\

In principle, it should be possible to get the same degree of polarization
for $e^-$ and $e^+$ beams. However some technical problems may arise and
it could be interesting to know the separate informations coming from
each lepton channel. Then we have decomposed our results into four classes :
{\it i)} only the $e^-$ beam is polarized;
{\it ii)} only the $e^+$ beam is polarized;
{\it iii)} both $e^-$ and $e^+$ beams are polarized, but we take into account
only the pure spin asymmetries $A$;
{\it iv)} both $e^-$ and $e^+$ beams are polarized, but now we take into account
all the possible spin and charge asymmetries ($A$ and $B$ types).
The distinction between the last two possibilities has a technical origin.
It comes from the fact that we expect stronger systematic errors for
the $B$-like asymmetries than for the $A$-like one's (see below).

\subsection{Experimental parameters}

\indent We present in this subsection, the parameters relevant to our
analysis. In order to preserve the clarity of the physical results presented
in the next section, we discuss here the effects due to some reasonable
variations of each parameter. Consequently, this part is rather technical
and could be skipped in a first reading.
Note also that the physical results only, have been given in
\cite{jmvhera97}.\\

\no $\bullet $ {\it \underline{Energy} :} 
\vspace{2.mm}

We have chosen $\sqrt{s} = 300\, GeV$ since it is the present value
for the HERA machine. The optimal choice $\sqrt{s} = 314\, GeV$ 
affects slightly our results, more precisely,
the forthcoming bounds have to be increased by roughly 2{\%}.  
This is in agreement with the scaling law \cite{Martyn87,Leike}:
$\Lambda_{lim} \sim (s.L)^{\frac{1}{4}}$, $L$ being the integrated
luminosity.\\

\no $\bullet $ {\it \underline{Integrated Luminosity} :} 
\vspace{2.mm}

We have considered the high luminosity option \cite{Felt,DeRoeck},
which corresponds to a "total" integrated luminosity
$L_{tot} = 1\, fb^{-1}$, for electrons ($L_-$) plus positrons ($L_+$).
When the beams are polarized, we have decomposed the respective
luminosities according to the spin configurations in order to keep
$L_{tot}$ constant, allowing us to make a realistic comparison
between the "discovery potentials" of the three distincts
0,1,2-spin analysis. Namely, it gives :
{\it i)} unpolarized : $L_- = L_+ = 1 \times 500\, pb^{-1} $;
{\it ii)} one-spin : $L_- = L_+ = 2 \times 250\, pb^{-1} $;
{\it iii)} two-spin : $L_- = L_+ = 4 \times 125\, pb^{-1} $.
In short, we take for example an integrated luminosity of $125\, pb^{-1}$
for the configuration $e^- (\lambda_e = +1)\, p (\lambda_p = +1)$
associated to the cross section $\sigma_-^{++}$, where we use
the notation $\sigma_t^{\lambda_e \lambda_p} \equiv 
{d\sigma_t}^{\lambda_e \lambda_p}/{dQ^2}$.

The choice $L_{tot} = 1\, fb^{-1}$ could be too high. Then we have
also considered the case where it is divided by a factor two.
It appears that the limits on $\Lambda$ decrease by about 15{\%}, which
is a little bit lower than the 
$\sqrt[4]{2}$
factor ($\sim 19 \% $)
indicated by the scaling law mentioned above.\\

\no $\bullet $ {\it \underline{Kinematic variables} :} 
\vspace{2.mm}

Concerning the $y$ variable (definition in appendix A), we take the usual
minimal cut \cite{HERAi} $y_{min} = 0.01$ (this choice has almost no
influence on our results). The maximal cut is fixed to $y_{max} = 0.95$,
a value which could be reached experimentally in the future at HERA
\cite{Martynpc}. If we use the present sensitivity \cite{HERAi} 
({\it i.e.} $y_{max} = 0.8$), all the limits decrease by {5 \%}.\\
Concerning the $Q^2$ resolution and the 
corresponding minimal cut we take the ones of
the ZEUS collaboration \cite{HERAi}: $\Delta Q^2/Q^2
= 34.3\,\%$ and $Q^2_{min} = 200\, GeV^2$. These precise choices have no visible
impact on the results.\\

\no $\bullet $ {\it \underline{Parton distribution functions} :} 
\vspace{2.mm}

For the unpolarized study we have used the three following unpolarized
parton distribution functions (pdf) sets : GRV \cite{GRV}, MRS \cite{MRS}
and BS \cite{BS}. The $\Lambda$ bounds are obtained with the GRV set,
but the other distributions give tiny differences.\\

For the polarized studies we have used again three polarized
pdf sets : BS \cite{BS}, GRSV \cite{GRSV} and GS96 \cite{GS96}.
However, the uncertainties on the polarized pdf's are presently
much larger than on the unpolarized one's. These uncertainties
will be strongly reduced in the near future, thanks to spin
asymmetries measurements at the RHIC-BNL polarized $pp$ collider
\cite{RSC},
for photon, jet and $W^\pm$ productions \cite{jsjmv}.

Consequently, we have calculated the cross sections and the asymmetries
with one pdf set, then we have obtained the bounds on $\Lambda$ 
assuming that the uncertainties due to the pdf will be weak
when the polarized HERA will be running. In other words,
we have assumed that anomalous effects due to new physics
will not be diluted by the uncertainties on pdf's.

The limits and the curves which will be presented, are obtained
with the GRSV set since it corresponds to the most conservative
attitude. Indeed, for this set the (valence) quarks are the
more weakly polarized, in comparison to GS96 and BS sets,
so it gives smaller spin effects and then smaller bounds.
Nevertheless, we can remark that the variations on the
$\Lambda_{eq}$ bounds from the different pdf sets ($\leq 3\%$)
are weaker in comparison to the variations on the
$\Lambda_{qq}$ bounds obtained from jet productions in hadron-hadron
collisions ($\leq 10\%$) \cite{ptjmvct,ptjmvctprd}.\\

\no $\bullet $ {\it \underline{Degree of beam polarization} :} 
\vspace{2.mm}

From an experimental point of view, the particle beams are never
fully polarized. We have to introduce a degree of polarization
$P$ for each beam. 
The first effect is to produce a "shift" between the magnitude
of the asymmetries defined theoretically ($P=1$) and measured
experimentally ($P\neq 1$). This induces a problem of definition
for the statistical error. We will comment on this difficulty in
a next paragraph, and in more details in appendix B.
A second effect is related to the uncertainty on the precise value of $P$,
which induces a systematic error for the asymmetries.

In the following we take for the degree of beam polarization of the
charged leptons ($P_e$) and of the protons ($P_p$), the values
which are in common use \cite{Felt,DeRoeck}: 
$P_{e^-}=P_{e^+}=P_p=70 \%$.
The choice $P_{all}=60 \%$ decreases the bounds for the one-spin
and the two-spin PV asymmetries by roughly 5$-$6{\%}, and the
ones for the two-spin PC asymmetries by roughly 8{\%}.
It comes from the fact that the statistical errors of the PV 
asymmetries are proportional to $1/P$ (see \cite{BRST} for $A_L$
and eq.(\ref{dasapv}) for $A_{LL}^{PV}$), 
and those of the PC asymmetries, are proportional to
$1/P^2$ (see eq.(\ref{aparr})).\\

\no $\bullet $ {\it \underline{Systematic errors on cross sections} :} 
\vspace{2.mm}

For the unpolarized cross sections, the systematic errors are rather
weak and of the order of $2-5 \%$ \cite{HERAi}\footnote{There are
some larger values in the low $Q^2$ region due to energy calibration
\cite{HERAi}, but we are not concerned with this domain of $Q^2$.}.
The main uncertainties come from luminosities measurements which
produce an unknown normalization factor $f$ \cite{Martyn87,HERAi}.
In the present analysis we do not take explicitly into account such a factor
$f$ and its related error $\Delta f (\sim 5 \%)$, like in Martyn's
analysis \cite{Martyn87,Martyn96}, but we rather follow a
procedure which is also convenient for the study of the asymmetries.
Namely, we take an explicit systematic error on the
involved observable, which is added in quadrature to the statistical
error of the same observable. The resulting total error is used in the
$\chi^2$ function (defined below).

For instance, we have chosen the systematic errors on the
unpolarized cross sections to be 
${\Delta \sigma_{syst}^{unpol}}/{\sigma^{unpol}} = 3\%$,
for the entire $Q^2$ domain. With this choice we recover to a good
accuracy the limits on $\Lambda$ presented in \cite{HERAi}.
However, this difference in the analysis strategy in comparison
to \cite{Martyn96}, gives some discrepancies on the $\Lambda$ bounds
for destructive interferences ($\epsilon = -1$), obtained from
the analysis of the unpolarized cross sections. 
For this special case, we find some limits 
which are roughly $15-20 \%$ higher than
the one's presented by Martyn in \cite{Martyn96}.
Nevertheless, we have to remark that this discrepancy in strategy
to take into account the systematic errors, disappears in the polarized
analysis since this normalisation uncertainty is irrelevant for an
asymmetry, which is a ratio of cross sections.\\

Concerning the polarized cross sections, there are some additional
systematic errors which are stronger in magnitude than for
unpolarized cross sections. They arise, on the one hand, from the 
uncertainties in the degree of polarization, and on the other hand,
from variations in counting rate due to time variations
in detector efficiencies, beam intensities, and crossing angles between 
the different spin configurations (see \cite{syst} for more details
on these systematics). These errors are added, so it is difficult
to have a global estimate. This is the reason why we prefer to
study spin asymmetries where these uncertainties largely
cancel in the difference of cross sections, which occurs in the numerator
of these asymmetries.

Without systematic errors, these polarized cross sections
are the most sensitive observables to the presence of new physics.
However, this "power of discovery" is quickly destroyed by
systematic errors. Then, it is irrelevant to give some constraints
on $\Lambda$ coming from the study of polarized cross sections if
we don't have any estimate for 
${\Delta \sigma_{syst}^{pol}}/{\sigma^{pol}}$.
In this spirit, we have estimated the values for
${\Delta \sigma_{syst}^{pol}}/{\sigma^{pol}}$ 
for which, the bounds
on $\Lambda$ obtained from the analysis of polarized
cross sections are at the same level than the ones obtained
from the analysis of spin asymmetries.
It appears that they have to be of the order of 10 \% (8 {\%})
in the one-spin (two-spin) case.\\

\no $\bullet $ {\it \underline{Systematic errors on the asymmetries} :} 
\vspace{2.mm}

 The use of asymmetries as observables in the analysis is motivated by 
the minimisation of systematic errors. On the one hand, the 
experimental systematic errors compensate in the numerator, 
as we have noted just above. On the other hand, the theoretical uncertainties 
in cross sections, coming from higher 
order corrections, which are expressed in general in term
of a K factor, partially cancel in the 
asymmetries \cite{BRST}, which are cross sections ratios.

   Unfortunately, these cancellations are not exact. Moreover, our 
numerical simulations are realized at the Leading Order.
Consequently, we have considered a global systematic error on the 
asymmetries of the order of 10\% :
\EQ
\frac{\Delta A_{syst}}{A}\; =\; \frac{\Delta B_{syst}}{B}\; =\; 10\, \% .
\eq
\no This value is the one expected at HERA in the polarized mode \cite{syst}.
\\

It appears that one can reduce this value by frequent reversals of 
spin orientations, a method which will be used at 
RHIC \cite{RSC}. At HERA, 
it is difficult to flip the lepton spin \cite{syst} and then,
this procedure is not very helpful to reduce
the systematic errors. Conversely, the spin flip of the proton 
beam will be helpful and then we expect the 
smallest systematic errors for the PC asymmetries,
defined below.\\
Concerning the charge asymmetries (B) we expect stronger systematics 
({\it i.e} $> $ 10\%) since we need data from
runs in the $e^-$ mode and in the $e^+$ mode, which cannot be 
done simultaneously.

   Finally, we have considered the impact of the variation of this
systematic error. It appears that the
choice $\Delta {\cal O}_{sys}/{\cal O}=20\%$ (${\cal O}=A$ or $B$)
decreases the bounds by roughly 10-15{\%}, the precise value depending on the 
involved asymmetry.\\

\no $\bullet $ {\it \underline{Statistical errors on the asymmetries} :} 
\vspace{2.mm}

When we take into account the degrees of polarization of the beams,
the definition of the statistical error of the corresponding asymmetry
may be ambiguous. For the spin asymmetries which have been 
studied/measured until now, like the PV one-spin asymmetry
$A_L\, (\equiv A_{LR})$ or the PC two-spin asymmetries $A_{LL}^{PC}$ and
$A_\parallel$, defined in deep inelastic 
scattering at low $Q^2$, 
the difficulty for the definition of $\Delta A_{stat}$
does not appear due to natural factorization ($A_L$, $A_{LL}^{PC}$) or relevant
assumptions ($A_\parallel$). (These asymmetries will be defined in the next section
or in appendix B).
In this paper, we consider a large set
of spin asymmetries where such natural factorizations do not
hold generally. Moreover, since we are concerned with the high $Q^2$
domain, we cannot make some assumptions, like parity conservation
relevant at low energy where, for instance, $A_\parallel$ measurements are
realized.\\
However, this problem is relatively technical and we prefer to insist
on the physical results that will be presented in the next sections,
then we give the definition of $\Delta A_{stat}$ in
appendix B.\\

\no $\bullet $ {\it \underline{$\chi^2 $ analysis} :} 
\vspace{2.mm}

The bounds on the scale $\Lambda$ are obtained with a $\chi^2$ 
analysis where the SM is the reference. The $\chi^2$ fonction is
defined by :
\EQ
\chi^2\; =\; \sum_{Q^2} \left( \frac{
{\cal O}^{SM+NP}(Q^2)\, -\, {\cal O}^{SM}(Q^2)}
{\Delta {\cal O}^{SM}(Q^2)} \right)^2 \; ,
\eq
\no where ${\cal O}$ is the involved observable, the indices $SM$ 
and $SM+NP$ refer respectively, to the values taken by 
$\cal O$ in the SM alone and in the SM with the New Physics (CI) contributions.
$\Delta {\cal O}$ is the quadratic sum of the statistical 
and systematic errors on ${\cal O}$. $\sum_{Q^2}$ corresponds to the 
sum over all the $Q^2$ bins defined in our analysis. 

We can also add to this function a sum over all the independent 
observables ${\cal O}$. However, it appears that it 
is not actually relevant in our analysis since we consider for the 
CI, only one ``individual'' model at once, whose contribution is, almost, 
mesurable in only one double polarized cross section.
This statement has to be reconsidered in the case of models involving
several ``chiralities'' ({\it i.e.} several $\eta_{ij}^\epsilon$).
\\

   The presence of a CI will induce an increase (from $0$) of the $\chi^2$  function. 
If we assume that no effects are detected,
we obtain a limit on $\Lambda$ at the 95\% Confidence Level (CL),
 if the $\chi^2$  increases, for this $\Lambda$, by $(1.96)^2$ 
\cite{PDG}\footnote{This remark is valid for some Gaussian statistics.
At very high $Q^2$, where the numbers of events are small, we have used
a Poisson statistics in agreement with \cite{PDG}.}. 
In the following all the bounds which are presented 
correspond to a 95\% CL.


\section{Definitions of the observables and $\Lambda_{eq}$ bounds}

\subsection{Unpolarized case}

\indent
The ``basic'' observables are the {\it two} unpolarized cross sections 
$\sigma_-$ and $\sigma_+$,
($\sigma_t \equiv {d\sigma_t}/{dQ^2}$).
With these two cross sections we can define the first (unpolarized) 
charge asymmetry, which was also considered in \cite{Ruckl84,Martyn87}:
\EQ
B_o\; =\; \frac{\sigma_- \, -\, \sigma_+}{\sigma_- \, +\, \sigma_+} \; .
\eq

   Using these observables, with the experimental parameters and the
$\chi^2$ analysis described above, we obtain
the bounds on $\Lambda$ presented in Table 3.1.
For each row, the relevant observable is mentioned, and $\epsilon = +1$
($\epsilon = -1$) corresponds to constructive (destructive) interferences.

\hspace{-0.8cm}
\begin{center}
\begin{tabular}{|c|c||c|c|c|c||c|c|c|c|}
\hline
$\Lambda$ (TeV)& Observable(s) & $\eta_{LL}^+$&$\eta_{RR}^+$&$\eta_{LR}^+$
&$\eta_{RL}^+$&$\eta_{LL}^-$&$\eta_{RR}^-$&$\eta_{LR}^-$&$\eta_{RL}^-$\\
\hline
\hline
$e^-$ only &$\sigma_-$& $6.2$ & $6.0$ & $2.6$ & $2.8$ & $5.4$ & $5.2$ & $1.8$ & $1.8$ \\
\hline
$e^+$ only &$\sigma_+$& $3.3$ & $3.4$ & $6.2$ & $6.0$ & $2.6$ & $2.7$ & $5.2$ & $5.0$ \\
\hline
$e^-$ and $e^+$  &$\sigma_- +\sigma_+$& $6.3$ & $6.1$ & $6.2$ & $6.0$ & $5.5$ 
& $5.3$ & $5.2$ & $5.0$ \\
\hline
$e^-$ and $e^+$ type $B$ &$B_o$& $3.1$ & $2.9$ & $4.7$ & $4.3$ & $3.1$ & $2.8$ & $3.8$ & $3.3$ \\
\hline
\end{tabular} 
\end{center}
\begin{center}
Table 3.1: Limits on $\Lambda$ at 95\% CL for the unpolarized case.
\end{center}

\no The conclusions from this first set of results are the followings :
\begin{itemize}
\item Depending on the lepton type, the cross sections are sensitive 
to certain chiralities : $\sigma_-$ ($\sigma_+$) tests the chiralities 
$LL^\pm ,\, RR^\pm $ ($LR^\pm ,\, RL^\pm $).
(This confirms our expectations from our analysis of the dominant terms 
performed in appendix A). 
If a non-standard effect is observed at HERA, the comparison of 
the two cross sections $\sigma_-$ and $\sigma_+$ 
allows the distinction of the two classes of chiralities 
($LL^\pm ,\, RR^\pm $) and ($LR^\pm ,\, RL^\pm $). 
Note that unpolarized cross
sections measurements are unable to discriminate within each 
classes, except if the destructive 
interferences pattern is actually seen, which seems to be difficult 
according to the experimental conditions (see next section).

\item For constructive interferences 
we are sensitive 
to $\Lambda (\eta_{ij}^+) \sim 6\, TeV$, and to $\Lambda(\eta_{ij}^-) \sim 5\, TeV$  
for destructive interferences.
See \cite{Martyn96} for comparison up to a factor 2 in integrated Luminosity
({\it i.e.} using the scaling law given in the preceding section
for a relevant comparison).

\item The compilation of $e^-$ and $e^+$ data does not increase 
significantly the limits on $\Lambda$ for a given chirality, 
but it allows the test of all the chiralities.

\item The measurement of the charge asymmetry $B_o$, does not give any 
complementary informations.
\end{itemize}

\no  The behavior of these updated results is analogous to the one's 
obtained some years ago in \cite{Ruckl84,Martyn87}.

\subsection{One-spin case}

\indent   The ``basic'' observables are the {\it four} single polarized cross 
sections $\sigma_-^-$, $\sigma_-^+$ and $\sigma_+^-$, $\sigma_+^+$,
($\sigma_t^{\lambda_e} \equiv {d\sigma_t^{\lambda_e}}/{dQ^2}$).
These four cross sections allow the definition of {\it two} ``pure'' 
spin asymmetries, which are PV ({\it i.e.}
non-zero for PV interactions) :
\EQ\label{defAL}
A_L(e^-)\; =\; \frac{\sigma^-_- \, -\, \sigma^+_-}
{\sigma^-_- \, +\, \sigma^+_-}
\;\;\;\;\;\;\;\;\;\;\;\;\; and \;\;\;\;\;\;\;\;\;\;\;\;\;
A_L(e^+)\; =\; \frac{\sigma^-_+ \, -\, \sigma^+_+}
{\sigma^-_+ \, +\, \sigma^+_+} \; ,
\eq

\no and {\it four} charge asymmetries, two by two independent :
\EQ\label{defB1a}
B_1^1\; =\; \frac{\sigma^-_- \, -\, \sigma^-_+}{\sigma^-_- \, +\, \sigma^-_+}
\;\;\;\;\;\;\;\;\;\;\;\;\; and \;\;\;\;\;\;\;\;\;\;\;\;\;
B_1^2\; =\; \frac{\sigma^+_- \, -\, \sigma^+_+}{\sigma^+_- \, +\, \sigma^+_+}\; ,
\eq
\EQ\label{defB1b}
B_1^3\; =\; \frac{\sigma^-_- \, -\, \sigma^+_+}{\sigma^-_- \, +\, \sigma^+_+}
\;\;\;\;\;\;\;\;\;\;\;\;\; and \;\;\;\;\;\;\;\;\;\;\;\;\;
B_1^4\; =\; \frac{\sigma^+_- \, -\, \sigma^-_+}{\sigma^+_- \, +\, \sigma^-_+}\; .
\eq
\no The lower index 1 indicates that only one beam is polarized. 
These asymmetries have been already defined in \cite{Ruckl84,Martyn87}.
In fact, we can construct two other charge asymmetries which involve
more than two independent cross sections,
but only the following one appears to have an interesting discovery potential :
\EQ
B_1^5\; =\; \frac{\sigma^-_-\, -\, \sigma^+_-\, + \,\sigma^-_+\, -\, \sigma^+_+}
{\sigma^-_-\, +\, \sigma^+_-\, +\, \sigma^-_+\, +\, \sigma^+_+} \; .
\eq

Concerning the $\Lambda$ bounds, from the $\chi^2$ analysis we 
have obtained the results presented in Table 3.2. 

\vspace*{-0.8cm}
\hspace*{-0.8cm}
\begin{center}
\begin{tabular}{|c|c||c|c|c|c||c|c|c|c|}
\hline
$\Lambda$ (TeV)& Observable(s) & $\eta_{LL}^+$&$\eta_{RR}^+$&$\eta_{LR}^+$
&$\eta_{RL}^+$&$\eta_{LL}^-$&$\eta_{RR}^-$&$\eta_{LR}^-$&$\eta_{RL}^-$\\
\hline
\hline
$e^-$ only &$A_L (e^-)$& $4.6$ & $5.6$ & $2.2$ & $3.0$ & $4.2$ & $5.2$ & $2.0$ & $2.6$ \\
\hline
$e^+$ only &$A_L (e^+)$& $2.8$ & $3.5$ & $4.4$ & $5.6$ & $2.0$ & $3.0$ & $4.0$ & $5.1$ \\
\hline
$e^-$ and $e^+$ type $A$ &$A_L (e^-) + A_L (e^+)$& $4.7$ & $5.7$ & $4.4$ & $5.6$ & $4.3$ 
& $5.3$ & $4.0$ & $5.1$ \\
\hline
$e^-$ and $e^+$ type $B$ &$B_1^{n}$& $4.3^5$ & $5.6^2$ & $5.6^2$ & $5.4^4$ & $4.0^5$ 
& $5.3^2$ & $5.1^2$ & $4.7^4$ \\
\hline
\end{tabular} 
\end{center}
\no Table 3.2: Limits on $\Lambda$ at 95\% CL for the one-spin case.
For the charge asymmetries 
$B_1^n$ (last row), the exponent $n$ which characterizes
the involved asymmetry, is also indicated for each corresponding
$\Lambda$ limit value. \\

\no Some comments are in order:
\begin{itemize}
\item The charge asymmetries $B_1^1$ and $B_1^3$ do not appear in the last row
of Table 3.2, which indicates that their sensitivity to CI are reduced in comparison
to the other asymmetries.

\item The $\Lambda$ bounds are much lower than in the unpolarized 
analysis, in particular for the $LL^\pm $ chirality.

\item We find a smaller difference (for $\Lambda$ limit) 
between the two conditions of constructive and destructive
interferences than in the unpolarized case.
%

\end{itemize}

   From a ``discovery potential'' point of view, we conclude that the 
analysis of one-spin asymmetries, defined when
lepton polarization is available, is less efficient than an 
unpolarized analysis. Such a conclusion has already been
given in \cite{Martyn87,Martyn91}.
However, if a new physics effect is detected, the analysis of 
one-spin asymmetries will be very useful to obtain some crucial
informations on the chirality structure of the new interaction 
\cite{Ruckl84,Martyn87}. It will be demonstrated in the next 
section for the two-spin case, since it is more powerful than the one-spin case.
Indeed, for models with some ``complex'' chiral structure ({\it i.e.}
involving several chiralities simultaneously), some cancellations
may occur, which reduce considerably the ``analysing potential''
of the one-spin asymmetries. \\
The interested reader could consult \cite{Ruckl84,Martyn87} to see 
the behavior of the different one-spin asymmetries.\\

\subsection{Two-spin case.}

\indent  The ``basic'' observables are the {\it eight} 
double polarized cross sections 
$\sigma_-^{--}, \sigma_-^{++}, \sigma_-^{-+}, \sigma_-^{+-} $
and $\sigma_+^{--}, \sigma_+^{++}, \sigma_+^{-+}, \sigma_+^{+-}$,
($\sigma_t^{\lambda_e \lambda_p} \equiv {d\sigma_t}^{\lambda_e 
\lambda_p}/{dQ^2}$).

We can construct {\it twelve} ``pure'' spin asymmetries involving two cross 
sections only. It turns out that the ones
which have the greatest ``analysing power'' are the two PV 
spin asymmetries :
\EQ\label{defALLPV}
A_{LL}^{PV}(e^-)\; =\; \frac{\sigma^{--}_- \, -\, \sigma^{++}_-}{\sigma^{--}_- \, 
+\, \sigma^{++}_-}
\;\;\;\;\;\;\;\;\;\;\;\;\; and \;\;\;\;\;\;\;\;\;\;\;\;\;
A_{LL}^{PV}(e^+)\; =\; \frac{\sigma^{--}_+ \, -\, \sigma^{++}_+}{\sigma^{--}_+ \, 
+\, \sigma^{++}_+}\; .
\eq

Of course, if the new interaction is PC, these two 
asymmetries are irrelevant. 
Therefore, it is interesting to study the
``discovery potential'' of the PC two-spin asymmetries. In addition, they 
could be important to disentangle the properties of a
new interaction having a more complex chiral stucture than the
individual models only. 
These PC asymmetries, all independent, are defined by :
\EQ\label{APC}
A_2^1\; =\; \frac{\sigma^{--}_- \, -\, \sigma^{-+}_-}{\sigma^{--}_- \, 
+\, \sigma^{-+}_-}
\;\;\;\;\;\;\;\;\;\;\;\;\; , \;\;\;\;\;\;\;\;\;\;\;\;\;
A_2^2\; =\; \frac{\sigma^{++}_- \, -\, \sigma^{+-}_-}{\sigma^{++}_- \, 
+\, \sigma^{+-}_-}\;\; ,
\eq
\EQ\label{APC2}
A_2^3\; =\; \frac{\sigma^{--}_+ \, -\, \sigma^{-+}_+}{\sigma^{--}_+ \, 
+\, \sigma^{-+}_+}
\;\;\;\;\;\;\;\;\;\;\;\;\; , \;\;\;\;\;\;\;\;\;\;\;\;\;
A_2^4\; =\; \frac{\sigma^{++}_+ \, -\, \sigma^{+-}_+}{\sigma^{++}_+ \, 
+\, \sigma^{+-}_+} \;\; .
\eq

   We see that they involve the cross sections with proton 
spin flip only. This could be particularly interesting to minimize 
systematic errors. Moreover, we notice that $A_2^2$ and $A_2^4$ 
are the spin asymmetries (usually called $A_\parallel$),
which are, in general, used to extract the polarized
structure function $g_1$. Three other spin asymmetries are of particular
interest :
\EQ
{\bar A}_{LL}^{PV}(e^-)\; =\; \frac{\sigma^{-+}_- \, 
-\, \sigma^{+-}_-}{\sigma^{-+}_- \, 
+\, \sigma^{+-}_-}
\;\; , \;\;
{\bar A}_{LL}^{PV}(e^+)\; =\; \frac{\sigma^{-+}_+ \, 
-\, \sigma^{+-}_+}{\sigma^{-+}_+ \, 
+\, \sigma^{+-}_+}
\;\; , \;\;
A_2^5\; =\; \frac{\sigma^{++}_- \, -\, \sigma^{-+}_-}{\sigma^{++}_- \, 
+\, \sigma^{-+}_-}\; .
\eq

\no Now, we want to introduce, among the huge number of charge asymmetries 
that can be defined, the ones which will have the largest
discovery potential in the present analysis :
\EQ\label{defB22}
B_2^{1}\; =\; \frac{\sigma^{-+}_- \, -\, \sigma^{--}_+}{\sigma^{-+}_- \, 
+\, \sigma^{--}_+} \;\;\;\;\;\;\;\; ,\;\;\;\;\;\;\;\;\;
B_2^2\; =\; \frac{\sigma_-^{++} \, -\, \sigma_+^{++}}{\sigma_-^{++} \, +\, 
\sigma_+^{++}} \;\; ,
\eq
\EQ
B_2^{3}\; =\; \frac{\sigma^{--}_- \, -\, \sigma^{++}_-\, +\, \sigma^{+-}_+ \, -\, 
\sigma^{-+}_+}{\sigma^{--}_- \, +\, \sigma^{++}_- \, +\, \sigma^{+-}_+ \, +\, 
\sigma^{-+}_+}
\;\; , \;\;
B_2^{4}\; =\; \frac{\sigma^{--}_- \, -\, \sigma^{++}_-\, +\, \sigma^{--}_+ \, -\, 
\sigma^{++}_+}{\sigma^{--}_- \, +\, \sigma^{++}_- \, +\, \sigma^{--}_+ \, +\, 
\sigma^{++}_+}
\; .
\eq



The bounds on $\Lambda$ from these different two-spin asymmetries are
presented in Table 3.3. We have to recall that these limits are strongly
dependent on the systematic errors assumed for the asymmetries, and we
refer to the discussion presented in section 2 concerning the 
expectations for $\Delta {\cal O}_{syst}/{\cal O}$ and the corresponding
dependence for the different sets of asymmetries.\\

\hspace*{-1.5 truecm}
\begin{tabular}{|c|c||c|c|c|c||c|c|c|c|}
\hline
$\Lambda$ (TeV)& Observable(s) & $\eta_{LL}^+$&$\eta_{RR}^+$&$\eta_{LR}^+$
&$\eta_{RL}^+$&$\eta_{LL}^-$&$\eta_{RR}^-$&$\eta_{LR}^-$&$\eta_{RL}^-$\\
\hline
\hline
$e^-$ only &$A_2^{n} {\scriptstyle (e^-)}$& $5.5^{\scriptstyle PV}$ & $6.0^{\scriptstyle PV}$ & $3.1^{5}$ & $3.4_{\scriptscriptstyle \overline{PV}}$ 
& $5.4^{\scriptstyle PV}$ & $5.8^{\scriptstyle PV}$ & $2.9^{5}$ & $3.1_{\scriptscriptstyle \overline{PV}}$ \\
\hline
$e^+$ only &$A_2^{n} {\scriptstyle (e^+)}$& $3.6^{4}$ & $3.9_{\scriptscriptstyle \overline{PV}}$ & $5.3^{\scriptstyle PV}$ & $6.0^{\scriptstyle PV}$ 
& $3.5^{4}$ & $3.6_{\scriptscriptstyle \overline{PV}}$ & $5.1^{\scriptstyle PV}$ & $5.8^{\scriptstyle PV}$ \\
\hline
$e^-$ and $e^+$ type $A$ &$A_2^{n} {\scriptstyle (e^-) +} A_2^{n} {\scriptstyle (e^+)}$& $5.7$ & $6.2$ 
& $5.4$ & $6.1$ & $5.6$ & $6.0$ & $5.2$ & $6.0$ \\
\hline
$e^-$ and $e^+$ type $B$ &$B_2^{n}$& $5.5^{4}$ & $6.1^{3}$ & $6.0^{2}$ & $6.0^{1}$ 
& $5.5^{4}$ & $5.9^{3}$ & $5.7^{2}$ & $5.5^{1}$ \\
\hline
$e^-$ and $e^+$ type $A$ &$A_{2}^{PC} $& $4.6^{1}$ & $4.4^{2}$ & $4.6^{4}$ 
& $4.3^{3}$ & $4.8^{1}$ & $4.6^{2}$ & $4.8^{4}$ & $4.5^{3}$ \\
\hline
\end{tabular} \\

\no Table 3.3: Limits on $\Lambda$ at 95\% CL for the two-spin case.
The label ``$\overline{PV}$'' corresponds
to the asymmetry $\bar{A}_{LL}^{PV}$.\\

\no The results have the following properties :
\begin{itemize}

\item For constructive interferences, the limits are at the same level
for $RR^+$ and $RL^+$ models and slightly lower for $LL^+$ and $LR^+$
models, than the ones obtained from the unpolarized analysis. For
destructive interferences the bounds are better. Roughly, we find that
the limits are comparable in magnitude to the unpolarized case,
and then, far better than for the one-spin case.
This result is drastically different than the one obtained for
quark-quark CI studies for jet production in
hadron-hadron collisions. In this case, we found that spin asymmetries
studies have a much greater discovery potential than unpolarized 
cross sections studies \cite{ptjmvctprd}. This fact is directly correlated
to the huge systematic errors associated to any unpolarized jet
cross sections.
\item The differences 
between the two conditions of constructive and destructive
interferences are reduced again, but we have still
$\Lambda (\eta_{ij}^+) > \Lambda (\eta_{ij}^-)$.

%
\item Informations on the chiral structure of the new interaction could be
obtained, and this, with a better sensitivity than for the one-spin case.
This analysis is the aim of the following section.
\end{itemize}


\section{Chiral structure analysis}


\indent In this section, we propose a strategy of analysis in order to obtain
some precise informations on the chiral structure of an eventual
new interaction, and on the nature of the interferences with
the SM ({\it i.e.} sign of $\epsilon$).\\
In a first step, for simplicity, we consider only the individual
models. Some remarks on the models with a more complicated
structure will be given in a following subsection. The curves are given
for $\Lambda = 4\; TeV$, in connection with the possible HERA
anomaly.\\

We have seen in the preceding section that the unpolarized
cross sections are sensitive to two classes of chiralities, depending
on the electric charge of the colliding lepton :\\
$\sigma_-$ is sensitive to ($LL^{\pm}$,$RR^{\pm}$) and
$\sigma_+$ to ($LR^{\pm}$,$RL^{\pm}$). Then,
it is necessary to run into the two channels 
($e^-p$ and $e^+p$)
in order to cover all the possible chiralities. 
Moreover, experimentally in order 
to obtain the value of $\epsilon$,
we need to identify the
destructive interference pattern, which could be feasible if
$\Lambda$ is small and the integrated luminosity very high. 
At HERA, with $L_{tot}=1\, fb^{-1}$, it needs $\Lambda < 3.5\, TeV$, a case
which is already excluded by Drell-Yan process analysis at the Tevatron
\cite{Bargerct2}.\\

Consequently, we cannot go further than the separation of the two
classes ($LL^{\pm}$,$RR^{\pm}$)-($LR^{\pm}$,$RL^{\pm}$) from
unpolarized cross sections measurements.
 The analysis of the different spin asymmetries allows a clear separation
of the different individual models. We will illustrate this fact with the
two-spin asymmetries.\\


\subsection{Two-spin case}

\begin{figure}[ht]
\begin{tabular}[t]{c c}
\centerline{\subfigureA{\psfig{file={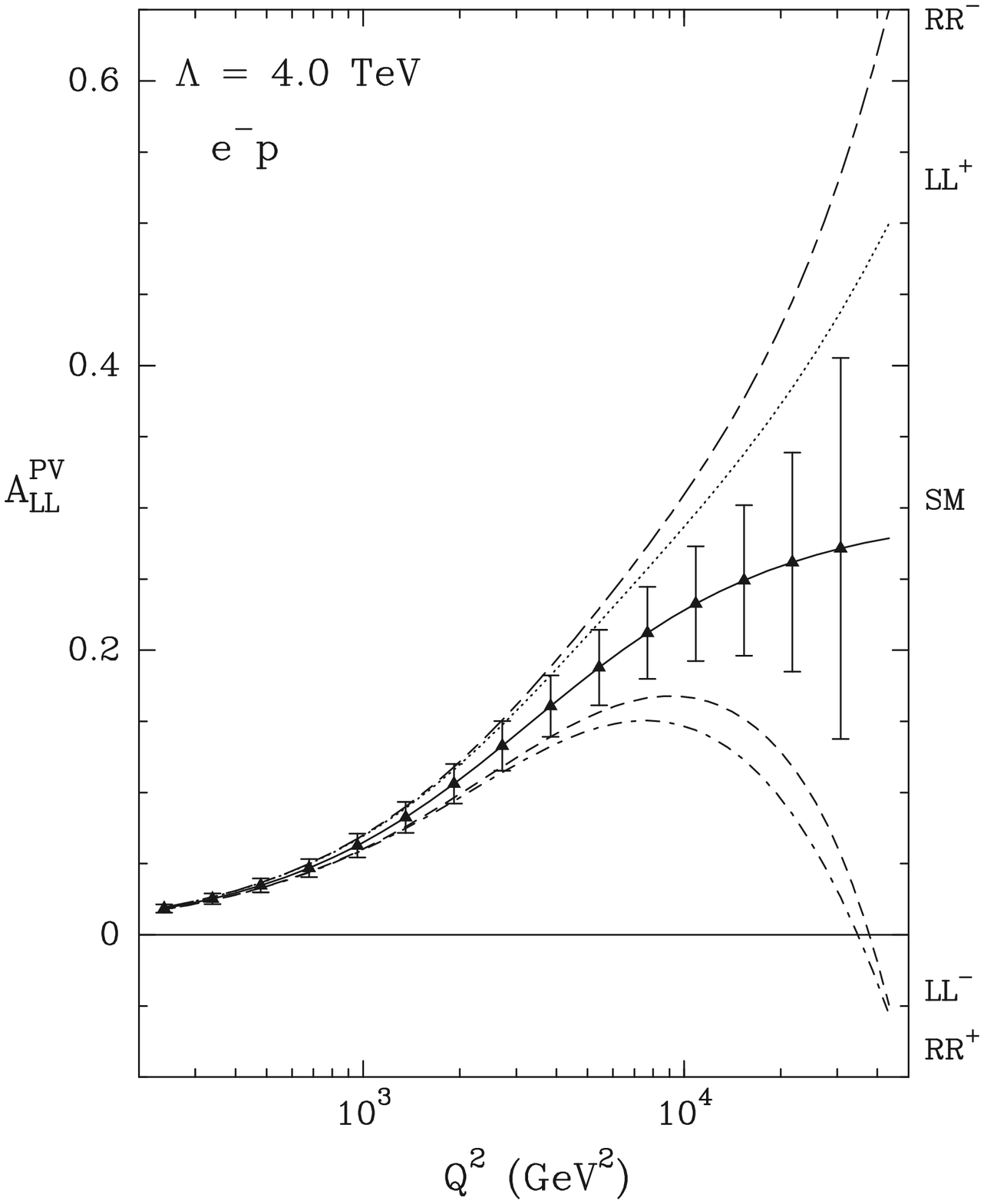},width=8truecm,height=10truecm}}
\subfigureA{\psfig{file={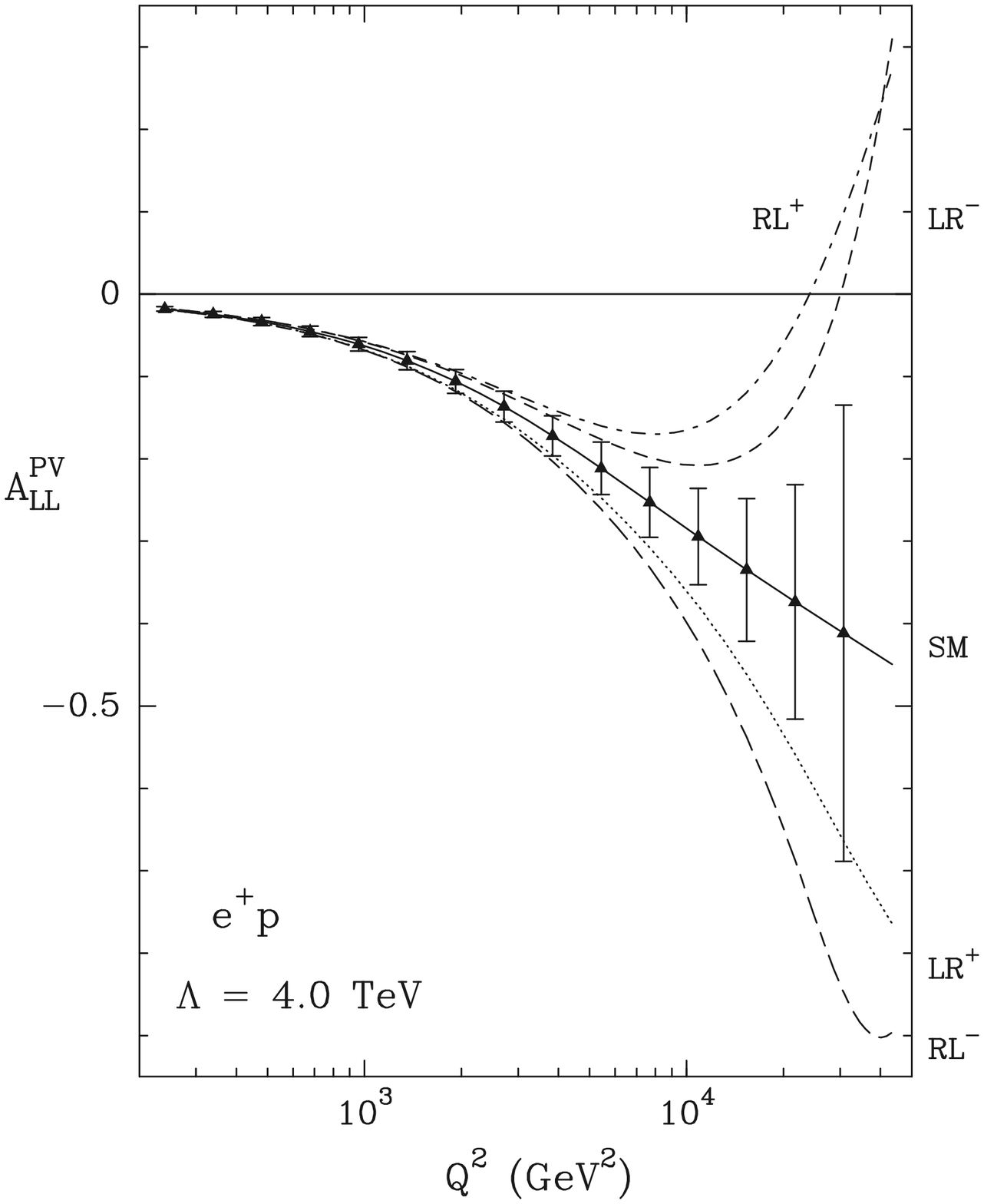},width=8truecm,height=10truecm}}}
\end{tabular} 
\vspace{-1cm}
\caption{\small Spin asymmetries $A_{LL}^{PV} (e^-)$ and 
$A_{LL}^{PV} (e^+)$. SM predictions (plain) and contact interactions 
for $\Lambda = 4\, TeV$, the relevant chiralities are given on the figure. 
Chiralities which are too close to the SM, are omitted.}
    \label{fig1} 
\end{figure}

The first spin asymmetries to consider are the PV one's
$A_{LL}^{PV} (e^-)$ and $A_{LL}^{PV} (e^+)$, which give the strongest
constraints on $\Lambda$. They are represented on Figure \ref{fig1}.

The behavior of these asymmetries is in perfect agreement with the 
expectations from the studies of their dominant terms, performed
in appendix A, which indicate for the numerator of the asymmetries :

\EQ
num [\, A_{LL}^{PV} (e^-) \, ] \; \simeq \; \epsilon\, 
K\, (+\eta + \eta\prime )\;\; u^+ \; ,
\eq
\EQ
num [\, A_{LL}^{PV} (e^+) \, ] \; \simeq \; \epsilon\, 
K\, (-\eta + \eta\prime )\;\; u^+ \; ,
\eq
\no where $u^+$ is the $u$ quark distribution for a quark helicity
parallel to that of the parent proton. 
$K={8 \pi \alpha^2}/{3Q^2 \Lambda^2}$
and $\eta$ and $\eta\prime$ characterize the chiral structure of the CI,
with the convention $LL\, (\eta = +1, \eta\prime = +1),\; RR\, (-1,-1),\; 
LR\, (+1,-1),\; RL\, (-1,+1)$.
Then, we deduce the following properties :
\begin{itemize}
\item Concerning $A_{LL}^{PV} (e^-)$, a deviation from the SM
expectation allows to separate the class ($LL^+,RR^-$) 
(positive deviation) from the class ($LL^-,RR^+$) (negative deviation).

\item Similarly, the study of $A_{LL}^{PV} (e^+)$ distinguishes ($LR^+,RL^-$)
from ($LR^-,RL^+$).

\item Then the comparison of $A_{LL}^{PV} (e^-)$ and $A_{LL}^{PV} (e^+)$
allows to pin down the origin of the effect coming from one of these 
four classes :
($LL^+,RR^-$) ($LL^-,RR^+$) ($LR^+,RL^-$) ($LR^-,RL^+$).
\end{itemize}

\begin{figure}[ht]
\vspace{-0.5cm}
    \centerline{\psfig{figure=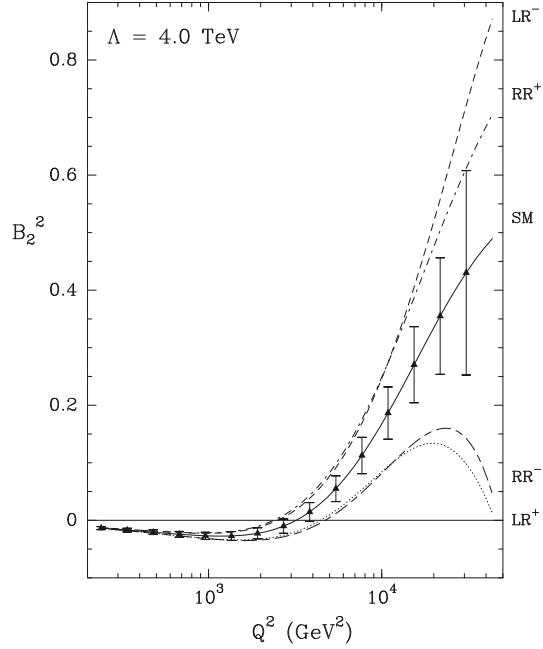,width=8cm}}
\vspace{-1cm}
    \caption[]{\small Same as Figure \ref{fig1} for $B_2^2$. }
    \label{fig2} 
\end{figure}

\no  The second step is to study the charge asymmetry $B_2^2$, which
is represented on Figure \ref{fig2}. We consider this asymmetry
because we have seen that it is the one which has
the best discovery potential for $LR^\pm$ models (under
the assumption ${\Delta B_{syst}}/{B}=10\,\%$).

\no Note that, as for $A_{LL}^{PV} (e^\pm)$, we can also obtain an approximate
formula for the numerator of $B_2^2$:
\EQ
num [\, B_2^2 \, ] \; \simeq \; -\, \epsilon\, 
K\, \eta (1- \eta\prime )\;\; u^+ \; .
\eq

\no The properties of this charge asymmetry are the followings :
\begin{itemize}
\item An effect on $B_2^2$ indicates the class
($LR^{\pm}$,$RR^{\pm}$). The direction of the deviation from SM
expectations allows the distinction between ($RR^+,LR^-$) for a 
positive deviation, and ($RR^-,LR^+$) for a negative one.

\item An absence of deviation on $B_2^2$ associated with a non-standard 
effect on an other observable ($A_{LL}^{PV} (e^\pm)$ in particular), 
reveals the presence of the class ($LL^{\pm},RL^{\pm}$).
\end{itemize}

\no The sum of the informations obtained from the three asymmetries
$A_{LL}^{PV} (e^-)$, $A_{LL}^{PV} (e^+)$ and $B_2^2$, allows us
to identify the chiral structure and the sign $\epsilon$ of
the new interaction (for the individual models).
This can be seen from Table 4.1 :\\

\begin{center}
\begin{tabular}{|c||c|c|c|c||c|c|c|c|}
\hline
 & $\eta_{LL}^+$&$\eta_{RR}^+$&$\eta_{LR}^+$
&$\eta_{RL}^+$&$\eta_{LL}^-$&$\eta_{RR}^-$&$\eta_{LR}^-$&$\eta_{RL}^-$\\
\hline
\hline
$A_{LL}^{PV} (e^-)$& $+$ & $-$ & $0$ & $0$ & $-$ & $+$ & $0$ & $0$ \\
\hline
$A_{LL}^{PV} (e^+)$& $0$ & $0$ & $-$ & $+$ & $0$ & $0$ & $+$ & $-$ \\
\hline
$B_2^2$            & $0$ & $+$ & $-$ & $0$ & $0$ & $-$ & $+$ & $0$ \\
\hline
\end{tabular} 
\\
\vspace{2.mm}

Table 4.1: ``Deviation signatures'' for the individual models.
\end{center}

\no In this table $+$, $-$, $0$ correspond to positive, negative and
``no-'' deviation from the SM. We see that each model
has a different ``deviation signature'', which indicates
a clear identification of the chiral structure for these
individual models. This identification could be realised up to a certain
scale, the "identification limit" $\Lambda_{id}$, which corresponds 
roughly to the
lower discovery limit obtained from 
$A_{LL}^{PV} (e^-)$, $A_{LL}^{PV} (e^+)$ and $B_2^2$. We find (see Table 3.3)
$\Lambda_{id} \sim 5.1\, TeV$.


\subsection{Models with chiral structure involving several individual terms}

In this subsection we do not attempt to realize a general study of these 
models with a more complicated chiral structure than the individual models.
Here, we just want to remark how the chiral
structure of some models with an extended structure, like the ones
advocated to explain the possible HERA anomaly,
could be constrained at a general level.\\

First, we have to note that, when several chiralities are present
simultaneously some cancellations may occur in the different asymmetries,
which can reduce drastically 
their analysing and discovery potentials.
However, in general, the three PV asymmetries 
$A_{LL}^{PV} (e^-)$, $A_{LL}^{PV} (e^+)$ and $B_2^2$ are sufficient
to obtain some valuable informations on the chiral structure of
the new interaction. This can be seen for the different models,
advocated to explain the possible HERA anomaly, which are the following.

\begin{itemize}
\item Barger et al.\cite{Bargerct1} have defined two models :
M1 : $\eta_{LR}^{\epsilon} = \eta_{RL}^{\epsilon}$ and M2 : 
$\eta_{LR}^{\epsilon} = \eta_{RL}^{-\epsilon}$ , the other $\eta_{ij}^\epsilon$
coefficients being set to zero.

\item The authors of ref.\cite{DiBart} defined three models :
VV : $\eta_{LL}^{\epsilon} = \eta_{RR}^{\epsilon} 
= \eta_{LR}^{\epsilon} = \eta_{RL}^{\epsilon}$,
AA : $\eta_{LL}^{\epsilon} = \eta_{RR}^{\epsilon} = \eta_{LR}^{-\epsilon} 
= \eta_{RL}^{-\epsilon}$
and VA : $\eta_{LL}^{-\epsilon} = \eta_{RR}^{\epsilon} 
= \eta_{LR}^{\epsilon} = \eta_{RL}^{-\epsilon}$.
(For completeness, 
we consider also the model AV : $\eta_{LL}^{-\epsilon} 
= \eta_{RR}^{\epsilon} = \eta_{LR}^{-\epsilon} = \eta_{RL}^{\epsilon}$)

\end{itemize}

\no Roughly, when we add the contributions of
the individual models for a given asymmetry, we find for M1, for instance,
that we will have no effect for $A_{LL}^{PV} (e^+)$ and 
a ``normal'' effect for $B_2^2$. We mean by "normal" effect, an effect of the
same magnitude than the one induces by an individual model only. Similarly,
we call a "double" effect, a deviation due to a certain model, which is twice
larger than the one due to an individual model. For instance,
for M2, we will have 
a double effect for $A_{LL}^{PV} (e^+)$ and 
a normal effect for $B_2^2$ (see Figures 1-2). 
Analogously for the other models,
we naively obtain the ``deviation signatures'' presented in Table 4.2.\\

\begin{center}
\begin{tabular}{|c||c|c|c|c|c|c|}
\hline
 & M1 & M2 & VV & AA & VA & AV \\
\hline
\hline
$A_{LL}^{PV} (e^-)$& $0$ & $0$ & $0$ & $0$ & $-2\epsilon$ & $-2\epsilon$ \\
\hline
$A_{LL}^{PV} (e^+)$& $0$ & $-2\epsilon$ & $0$ & $0$ & $-2\epsilon$ & $2\epsilon$\\
\hline
$B_2^2$  & $-\epsilon$ & $-\epsilon$ & $0$ & $2\epsilon$ & $0$ & $2\epsilon$ \\
\hline
\end{tabular} 
\\
\vspace{2.mm}

Table 4.2: ``Deviation signatures'' for the models with several terms.
\end{center}

\begin{figure}
\begin{tabular}[t]{c c}
\centerline{\subfigureA{\psfig{file={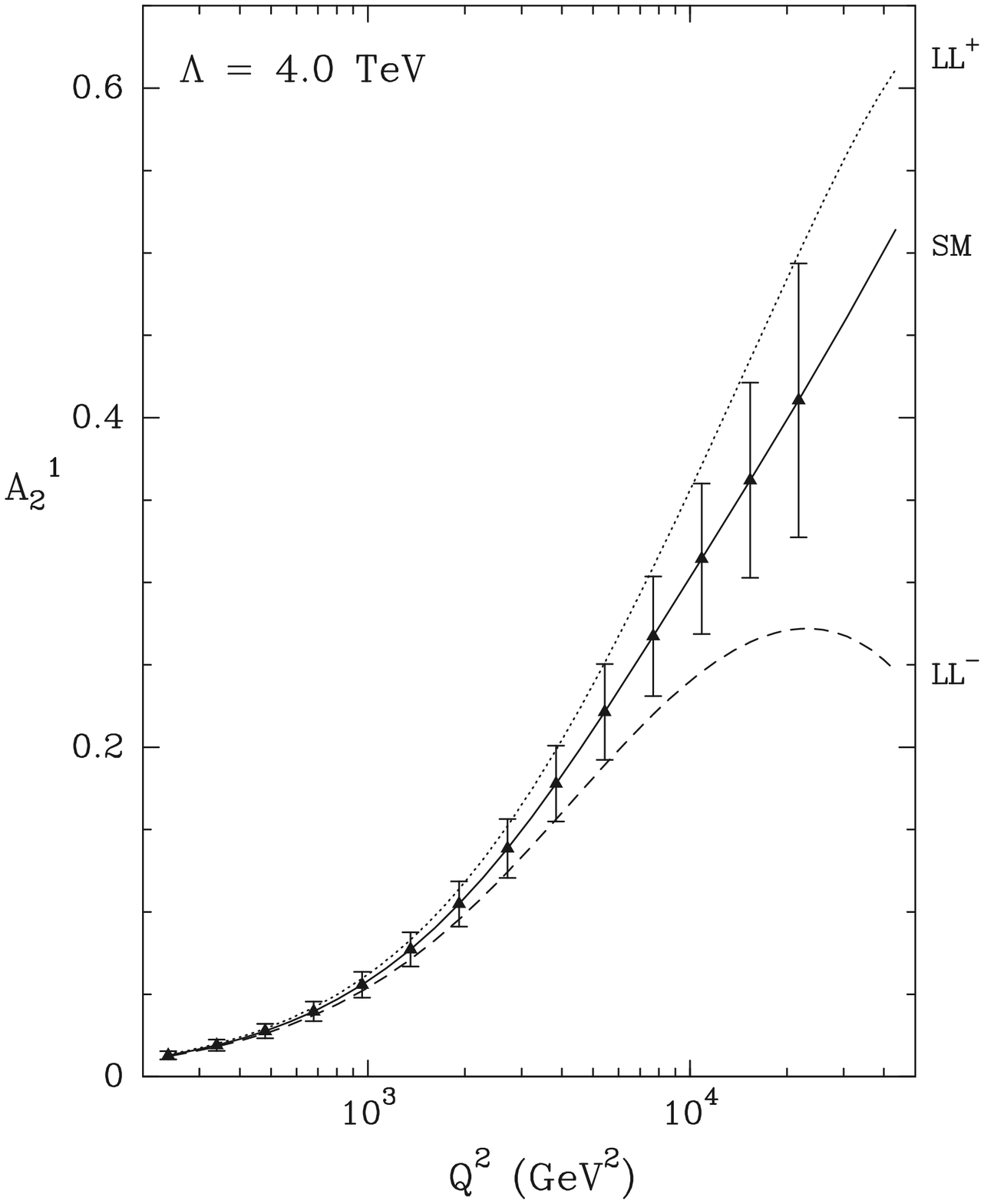},width=8truecm,height=10truecm}}\subfigureA{\psfig{file={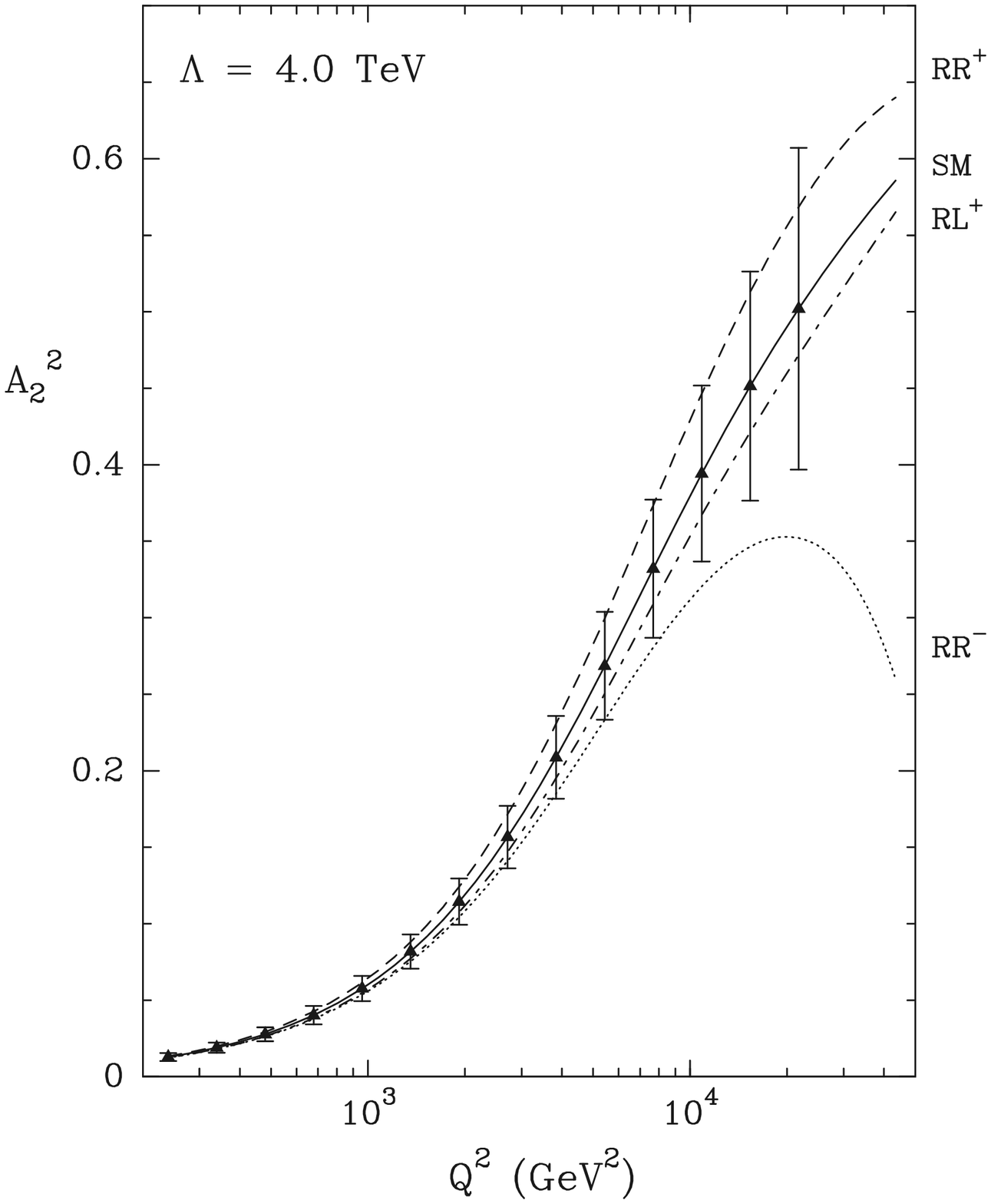},width=8truecm,height=10truecm}}}\\
\centerline{\subfigureA{\psfig{file={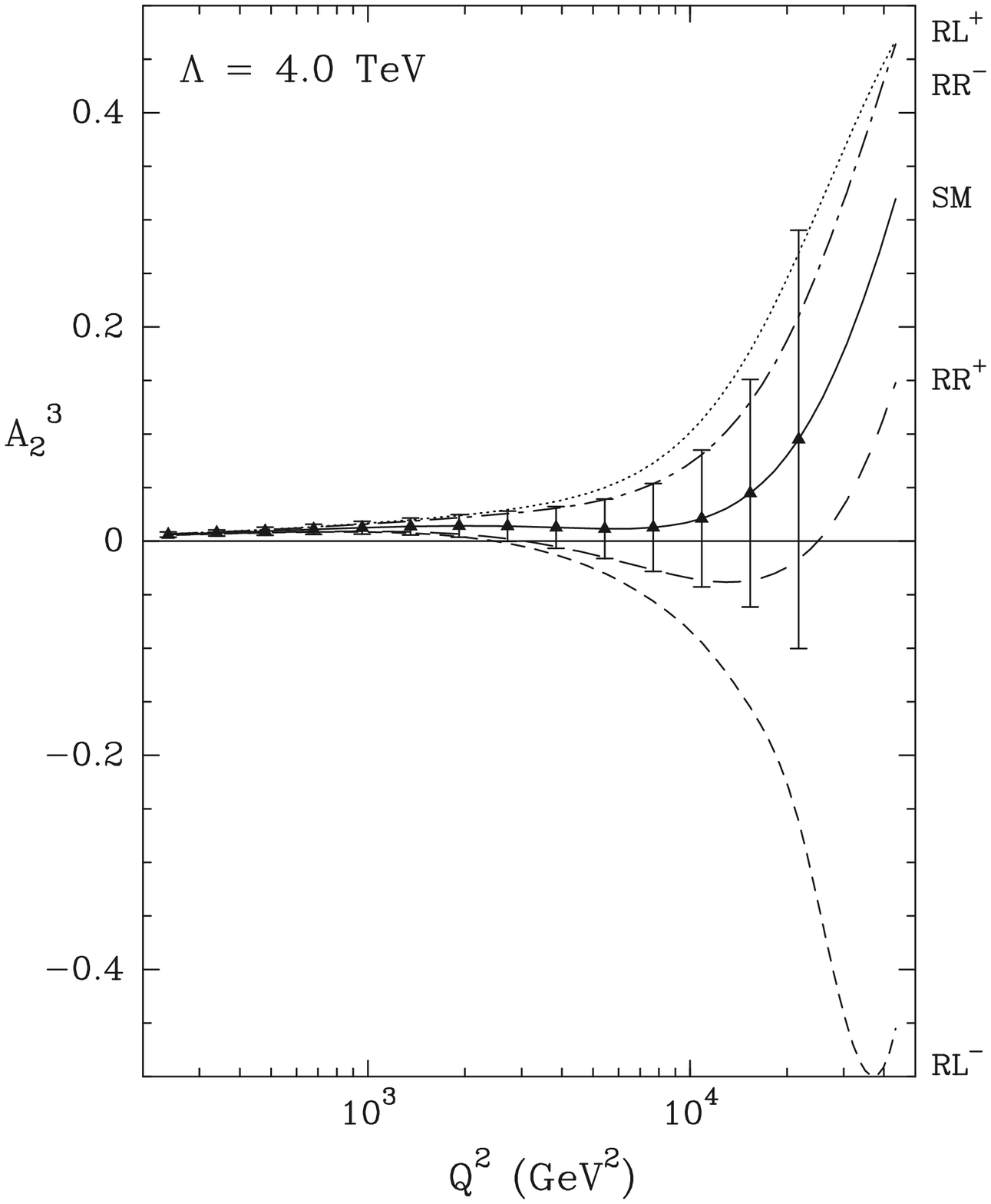},width=8truecm,height=10truecm}}\subfigureA{\psfig{file={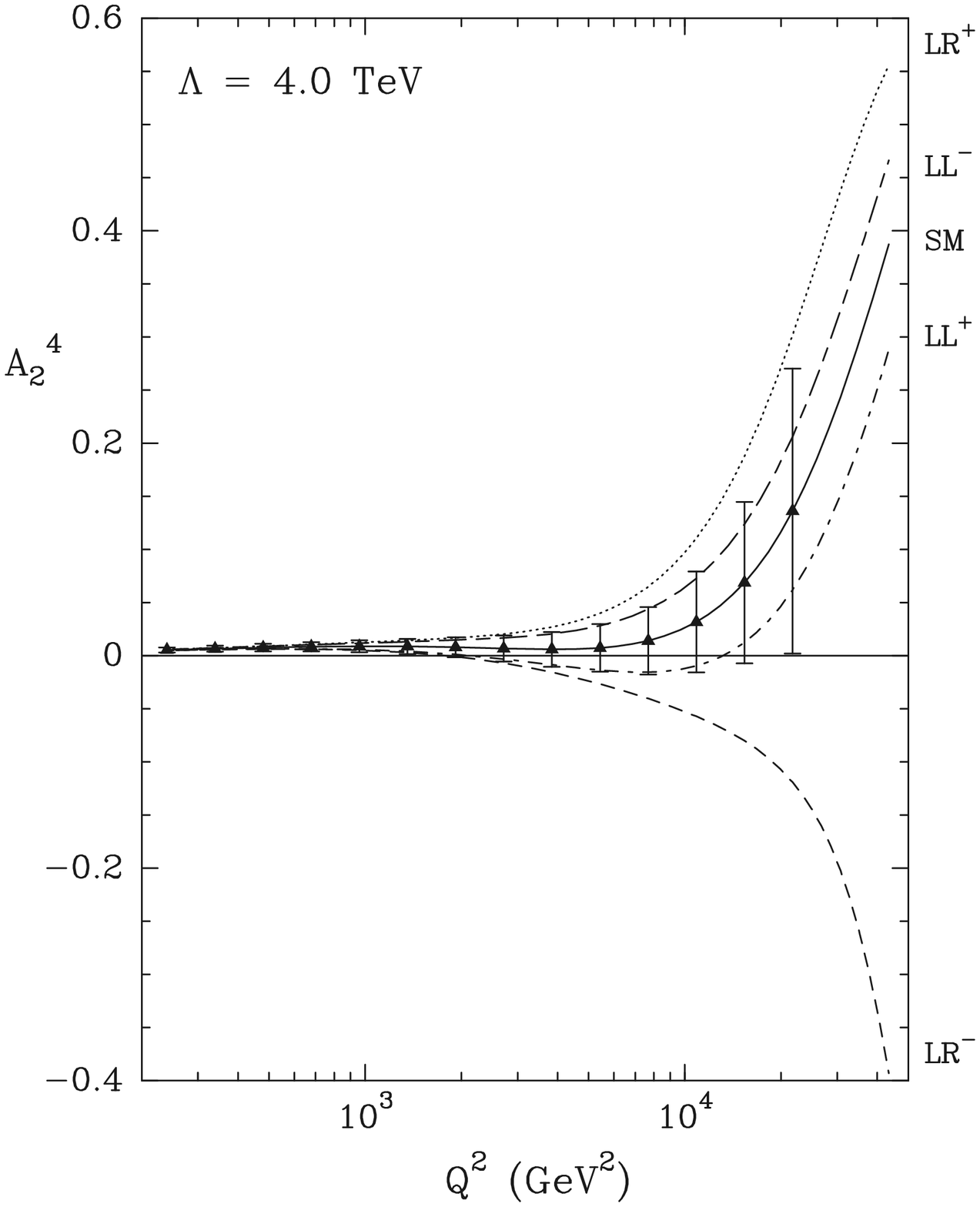},width=8truecm,height=10truecm}}}
\end{tabular} 
\caption{Same as Figure \ref{fig1} for Parity Conserving Asymmetries.}
\label{fig3}
\end{figure}

\no We find that, in general, the ``deviation signatures'' are distinct
between these models and also distinct from the one's of the individual models
given in Table 4.1. However, some ambiguities remain with 
several models. For instance, it could be difficult to
distinguish between the M2 model and the $\eta_{LR}^\epsilon$ term alone.
Moreover, the VV model does not give any deviation, since it conserves 
parity\footnote{
In fact, there are some small effects due to Z-CI interferences,
sizeable in the high $Q^2$ domain, if $\Lambda$ is not too large.}.
It is for such distinctions that the studies of the different PC
spin asymmetries are needed.\\
Indeed (see appendix A), it appears that the cross sections defined with identical
helicities ({\it i.e.} $\sigma_t^{\lambda\lambda}$ with
$\lambda =\lambda_e=\lambda_p$) are mainly
sensitive to one chirality only. 
Conversely, the cross sections $\sigma_t^{\lambda -\lambda}$
are almost unsensitive to new physics.
Then if we construct some spin asymmetries defined with these two types
of cross sections, we will obtain some asymmetries which are mainly
sensitive to one chirality only.
Such asymmetries correspond to the PC asymmetries defined
in section 3 (eq.(\ref{APC}-\ref{APC2})). We recall that we 
have found that they are less
sensitive to new physics than the PV asymmetries (see Table 3.3). These four
PC asymmetries are represented for the individual models on Figure
\ref{fig3}.\\

In conclusion, we deduce from this rough analysis, that in general,
the PV spin asymmetries are sufficient to obtain some crucial
informations on the chiral structure of the new interaction, but
it is necessary to study the PC asymmetries in order to disentangle
some hidden effects.


\subsection{Remarks on the one-spin asymmetries}

\indent The procedure described above could be easily adapted to the 
one-spin asymmetries. If we remember that the one-spin cross sections are
given in terms of the two-spin cross sections,
we see immediatly that the numerators of 
$A_{LL}^{PV} (e^-)$, $A_{LL}^{PV} (e^+)$ and $B_2^2$ are contained
in the numerator of the one-spin asymmetries 
$A_{L} (e^-)$, $A_{L} (e^+)$ and $B_1^2$, respectively.
Then, the study of these three one-spin asymmetries is roughly equivalent
to that of the three two-spin asymmetries. However, several
facts make this one-spin analysis less significant:

\begin{itemize}

\item The presence in the numerators of the $\sigma_t^{\lambda\, - \lambda}$
cross sections decreases the discovery potential, in the sense that
the ``informations'' contained in the numerators are diluted, since these
cross sections are weakly sensitive to new physics. This can be understood
by the examination of 
the ratio of the dominant contributions of the one-spin to two-spin
cross sections, given by (see appendix A) :
\EQ
\frac{\sigma_t^{\lambda_e}}{\sigma_t^{\lambda_e\lambda_p}}\; \simeq\; \frac{u/2}{u^+} 
\; =\; 1 - \frac{\Delta u}{2 u^+}\;\; < 1
\;\;\;\;\;\;\;\;\;\;\; since \;\; \Delta u > 0 \; .
\eq
\no This is confirmed by the results presented in Tables 3.2 and 3.3,
where we see that the one-spin asymmetries are sensitive to CI for 
some energy scales smaller by roughly
$1\; TeV$ in comparison to the sensitivity of the two-spin asymmetries.

\item We have at hand a smaller number of spin and charge asymmetries
in comparison to the two-spin case. In particular, we {\it cannot}
define the Parity Conserving asymmetries $A_2^{1-4}$. Then, if
the chiral structure of the new interaction is complex, and if some
cancellations occur between the different individual terms, we
can lose some important informations on
this chiral structure\footnote{One can argue that a fitting procedure
to the data, combining unpolarized and single polarized observables,
will constrain all the different chiralities \cite{Martynzprim}.
However, the arguments presented here, indicate that the fitting
procedure in the one-spin case will be less sensitive to new physics
than in the two-spin case, which will results in weaker constraints.} !
\end{itemize}


\section{Conclusions}

\begin{itemize}

\item The HERA collider, with a high integrated luminosity 
($L_{tot}=1\, fb^{-1}$), could
give some strong bounds on the energy scale of a possible new 
CI. For constructive interferences, the scale $\Lambda$
is of the order of $6\; TeV$, and for destructive interferences
we get $\Lambda \sim 5\; TeV$. The availability of 
polarized leptons and protons
beams will not increase significantly these bounds, except for
destructive interferences.
When only lepton polarization is available the sensitivity is
strongly reduced, the limits decreasing by roughly $1\; TeV$.

\item Contrarily to the analysis that can be performed for
unpolarized collisions, the studies of several spin and charge
asymmetries defined in a polarized context, can give some
crucial informations on the chiral structure of the new
interaction. The nature, constructive or destructive, of the 
interferences with photon exchange, can be also disentangled.
It appears that the asymmetries defined for double polarized
collisions are more well-suited for this chiral structure analysis,
than the asymmetries defined with lepton polarization only.
In any case, the availability of electron plus positron beams
is mandatory, in order to cover all the possible chiralities. 

\item This analysis has assumed an universality for $u$ and $d$
quarks contributions ($\eta_{ij}^u=\eta_{ij}^d$). 
However, in a proton, the $u$-quark
distribution is dominant. Consequently, the results
presented here essentially constrain the presence of a new interaction in the
electron-up-quark sector. To constrain an electron-down-quark
new interaction, the isospin symmetry indicates that the protons have to
be replaced by neutrons, so one should use electron-$He^3$ collisions,
an option which is also under consideration at HERA \cite{DeRoeck}.

\item In this paper we concentrate on the presence of some new CI.
However, there are some equivalences between such
CI and the exchange of some new Leptoquarks or
{\rp }-squarks. It is well-known \cite{Alt1} that such bosons
have to be chiral ({\it i.e.} with PV couplings) to
evade the present experimental constraints.
Then, the general results on the chiral structure analysis presented here, could be
applied to these new bosons if they have an accessible mass at HERA,
since, in this case, they will induce some important effects
in the different PV spin asymmetries. 
These questions will be analyzed in a forthcoming paper \cite{etjmv}.

\end{itemize}


\vspace*{3cm}

\no {\bf Acknowledgments}

\m \no

I am indebted to P. Taxil and J. Soffer for invaluable discussions,
and to A. De Roeck, H.-U. Martyn, B. Kamal, C. Benchouk,
G. Bunce, M.J. Tannenbaum
and J. Feltesse for very helpful informations and discussions. 
In particular, I am gratefull to J. Soffer, M.J. Tannenbaum and all
the active members of the "RHIC Spin Discussion
Team" at the RIKEN-BNL Research Center,
for encouraging me to write the appendix B.
I would like to 
thank M. Karliner who has initiated this work (a long time ago !). 
I wish to acknowledge Professors T.D. Lee and A. Baltz for their invitation
at the RIKEN-BNL Research Center, where part of this work has been done.
Finally, I thank P. Esposito for her kind hospitality at BNL.

\vspace{2.cm}

\newpage

\appendix
\section{ Formulas}

\appendixA{}

We present in this appendix, the set of formulas necessary to calculate
the double polarized cross sections, the spin and charge asymmetries
involved in the present analysis and, finally, the dominant contributions
coming from the presence of a new CI.

\subsection{Cross sections}

\no $\bullet $ {\it \underline{Process}}\\

\no The single polarized cross sections are given in \cite{Ruckl84,Cash}.
Here we give the cross sections in the ($s$, $t$, $u$) notations. \\
\no The collisions between charged leptons and protons, in the neutral
current channel, correspond to the process :
${\vec{e}}^{\,\pm}\; \vec{p}\; \longrightarrow\; e^{\pm}\; X$,
whose cross section is given by :

\EQ\label{defsec}
{\sigma_t}^{\lambda_e \lambda_p}\; \equiv\;
{\frac{d\sigma_t}{dQ^2}}^{\lambda_e \lambda_p}\; =\; \sum_q \int^{x_{max}}_{x_{min}} dx\,
\sum_{\lambda_q}
{\frac{d\hat{\sigma}_t}{d\hat{t}}}^{\lambda_e \lambda_q}\, q^{\lambda_q}_{\lambda_p}
(x,Q^2)
\eq

\no where $\lambda_e$, $\lambda_p$ and $\lambda_q$ are the helicities
of the charged lepton, proton and parton (quark or antiquark), respectively.
The label $t = \pm$ of $\sigma_t$  corresponds to the electric charge
of the colliding lepton.
$\sum_q$ represent the sum over all the quark and antiquark flavors
present inside the proton.
Concerning the kinematical variables,
let $P_N$, $P_l$ and $P_{l'}$ be the momentum of the proton, the incoming
lepton and the scattered lepton. If $P_q= x\, P_N$ is the momentum of the parton carrying the momentum fraction
$x$ of the nucleon, the subprocess variables ${\hat s}$, ${\hat t}$ 
and ${\hat u}$ are given by :
\EQ\label{defsh}
{\hat s} = (P_q + P_l)^2 \simeq  +2\, P_q P_l = x\, 2\, P_N P_l = x\, s\;\;\;\;\;\;\;\; ,
\eq
\EQ
{\hat t} = (P_l - P_{l'})^2  = t = -Q^2\;\;\;\;\;\;\;\;\;\;\;\;\;\;\;\;\;\;\;\;\;\;\;
\;\;\;\;\;\;\;\;\;\; ,
\eq
\EQ\label{defuh}
{\hat u} = (P_q - P_{l'})^2  \simeq  -2\, P_q P_{l'} = x\, u = -x(1-y)\, s \; ,
\eq
\no where the usual variable $y$ is defined by $y=P_N(P_l-P_{l'})/P_NP_l\; =
Q^2/xs$. The boundary condition $x\leq  1$ implies $y\geq {Q^2}/{s}$.
However, experimentally the $y$ variable is constrained in the range \cite{HERAi}:
$y_{min}^{exp} = 0.01$  and  $y_{max}^{exp} = 0.8$, but one is
hoping to reach in the future \cite{Martynpc} : $y_{max}^{exp} = 0.95$.
The integration limits are :
$x_{min}={Q^2}/{s\, y_{max}},\; 
x_{max}={Q^2}/{s\, y_{min}}$
with $y_{min} = max (y_{min}^{exp}, {Q^2}/{s}),\; y_{max} = y_{max}^{exp}$.
In the high $Q^2$ region we have roughly : $x_{min} \simeq
{Q^2}/{s},\; x_{max}\simeq 1$.\\

\no We denote by
$q^{\lambda_q}_{\lambda_p}(x,Q^2)$ the parton distribution for
the parton $q$ inside a proton
of helicity $\lambda_p$, with momentum fraction $x$ and helicity $\lambda_q$, 
at the energy scale $Q^2$. 
These distributions are related to the
parallel and anti-parallel distributions by : $q_+\, =\, q_+^+ \, 
=\, q_-^- \;\; , \;\;
q_-\, =\, q_+^- \, =\, q_-^+ $, which are related to the usual
unpolarized and polarized parton distributions by :
$q\, = \, q_+ + q_-$ and $\Delta q\, = \, q_+ - q_-$.\\


\no $\bullet $ {\it \underline{Subprocesses}}\\

\no Using the notations of \cite{BGS}, the cross section of the elementary
subprocess ${\vec{e}} \,\vec{q} \rightarrow e\, q$ is given by :
\EQ
{\frac{d\hat{\sigma}_t}{d\hat{t}}}^{\lambda_e \lambda_q}\; =
\; \frac{\pi}{{\hat s}^2}\;
\sum_{\alpha , \beta}\; T_{\alpha , \beta}^{\lambda_e \lambda_q}(e^t,q) \; ,
\eq

\no where $T_{\alpha , \beta}^{\lambda_e \lambda_q}(e^t,q)$ is
the square matrix element for $\alpha$ and $\beta$ boson exchange,
or the exchange process replaced by a CI; $q$
being a quark or an antiquark.
The $T_{\alpha , \beta}^{\lambda_e \lambda_q}(e^t,q)$ are given
below for $\alpha , \beta = \gamma ,\, Z,\, CI$. We have
omitted the hat symbol of the variables ${\hat s}$, ${\hat t}$ 
and ${\hat u}$, for clarity.\\

\no {\it\bf Subprocess}\hu {\bf $e^-\; q\; \longrightarrow \; e^-\; q$} :
\EQA
T_{\gamma \gamma} &=& 2\, e_q^2\, \frac{\alpha^2}{t^2}\, [\,(1+\lambda_e \lambda_q )
\, s^2 + (1-\lambda_e \lambda_q )\, u^2\, ]\; ,
\\ 
&& \nonumber \\
T_{Z Z} &=&  \frac{\alpha_Z^2}{t_Z^2}\, [ \,\left(C_{eL}^2 C_{qL}^2
(1-\lambda_e)(1-\lambda_q) + C_{eR}^2 C_{qR}^2 (1+\lambda_e)(1+\lambda_q) \right)
\, s^2 \nonumber \\
&&+ \left(C_{eL}^2 C_{qR}^2
(1-\lambda_e)(1+\lambda_q) + C_{eR}^2 C_{qL}^2 (1+\lambda_e)(1-\lambda_q) \right)
\, u^2\, ]\; ,
\\
&& \nonumber \\
T_{\gamma\, Z} &=& -\, 2\, e_q\, \frac{\alpha\alpha_Z}{tt_Z}\, [ \,\left(C_{eL} C_{qL}
(1-\lambda_e)(1-\lambda_q) + C_{eR} C_{qR} (1+\lambda_e)(1+\lambda_q) \right)
\, s^2 \nonumber \\
&&+ \left(C_{eL} C_{qR}
(1-\lambda_e)(1+\lambda_q) + C_{eR} C_{qL} (1+\lambda_e)(1-\lambda_q) \right)
\, u^2\, ]\; ,
\\
&& \nonumber \\
T_{CI CI} &=&  \frac{1}{2 \Lambda^4}\, [ \,\left((1+\eta {\eta}')
(1+\lambda_e\lambda_q) + (-\eta -{\eta}') (\lambda_e+\lambda_q) \right)
\, s^2 \nonumber \\
&&+ \left((1-\eta {\eta}')
(1-\lambda_e\lambda_q) + (\eta -{\eta}') (-\lambda_e+\lambda_q) \right)
\, u^2\, ]\; ,
\\
&& \nonumber \\ \label{TgaCI}
T_{\gamma\, CI} &=& -\,\epsilon\, e_q\, \frac{\alpha}{t \Lambda^2}\, 
[ \,\left((1+\eta {\eta}')
(1+\lambda_e\lambda_q) + (-\eta -{\eta}') (\lambda_e+\lambda_q) \right)
\, s^2 \nonumber \\
&&+ \left((1-\eta {\eta}')
(1-\lambda_e\lambda_q) + (\eta -{\eta}') (-\lambda_e+\lambda_q) \right)
\, u^2\, ]\; ,
\\
&& \nonumber \\
T_{Z\, CI} &=& \epsilon\, \frac{\alpha_Z}{2 t_Z \Lambda^2}\, [ \,(\,
C_{eL} C_{qL} (1+\eta) (1+{\eta}') (1-\lambda_e) (1-\lambda_q) 
\nonumber \\
&&+ C_{eR} C_{qR} (1-\eta) (1-{\eta}') (1+\lambda_e) (1+\lambda_q)\, )
\, s^2 \nonumber \\
&&+(\,
C_{eL} C_{qR} (1+\eta) (1-{\eta}') (1-\lambda_e) (1+\lambda_q) 
\nonumber \\
&&+ C_{eR} C_{qL} (1-\eta) (1+{\eta}') (1+\lambda_e) (1-\lambda_q)\, )
\, u^2\, ]\; ,
\eqa

\no where $\alpha$ is the electromagnetic coupling,
$\alpha_Z = \alpha/\sin^2 \theta_W \cos^2 \theta_W$, $t_{Z} = 
t-M^2_{Z}$. $C_{fL}$ and $C_{fR}$ are the usual Left-handed and Right-handed
couplings of the $Z$ to the fermion $f$, given by
$C_{fL}=I^f_3 - e_f \sin^2 \theta_W $, $C_{fR}=- e_f \sin^2 \theta_W $
with $I_3^f = \pm 1/2$.
The parameters $\eta$ and ${\eta}'$ characterize the
chiral structure of the CI with the conventions :
$LL\, (\eta = +1, \eta\prime = +1),
\; RR\, (-1,-1),\; LR\, (+1,-1),\; RL\, (-1,+1)$.\\

\no {\it\bf Subprocess}\hu {\bf $e^+\; q\; \longrightarrow \; e^+\; q$} :\\

The squared matrix elements $T_{\alpha ,\beta}$ are obtained from the six
preceding equations with the following changes :
$s \longleftrightarrow u$ and $\lambda_e \longleftrightarrow -\lambda_e$
(equivalent to $C_{eL} \longleftrightarrow C_{eR}$ and
$\eta \longleftrightarrow -\eta$).\\

\no {\it\bf Subprocess}\hu {\bf $e^{\pm}\; {\bar q}\; \longrightarrow \; 
e^{\pm}\; {\bar q}$} :\\

The $T_{\alpha ,\beta}$ are obtained from the ones for $e^{\pm} q$
scattering after the transformations :
$s \longleftrightarrow u$ and $\lambda_q \longleftrightarrow -\lambda_q$.
\\

\no The dominant term being $T_{\gamma\, CI}$ (see below), 
$\epsilon = +1$ corresponds to constructive interferences since
$t < 0$ and that in a proton the $u$-quark, with $e_u > 0$, is dominant.\\

\subsection{Dominant terms}

Some general arguments on the dominant effects, due to the presence of a CI, 
will allow us to obtain 
some simple relations for the cross sections, which in turn will give
the possibility to predict the behavior of the different asymmetries.
For simplicity, we realize these studies with the double differential
cross sections ${d\sigma^{\lambda_e\lambda_p}}/{dx dQ^2}$, the integration
over $x$ does not change the results.\\
The important point is that we study the events with the highest energies,
namely with some high $Q^2$, $x$ and $y$. Then we can make the following
reasonable hypothesis : 
\begin{itemize}
\item $H1$ \r ${\hat s}^2 \gg {\hat u}^2$, 
\item $H2$ \r $u(x,Q^2) > d(x,Q^2)$, 
\item $H3$ \r $\Delta u(x,Q^2) > 0$ (and increasing with $x$, for high $x$), 
\item $H4$ \r ${\bar q}(x,Q^2) \ll q(x,Q^2)$.
\end{itemize}

\no With $H4$ the process cross section corresponding to eq.(\ref{defsec}) becomes :
\EQ
{\sigma}^{\lambda_e \lambda_p}_t\; \equiv\;
{\frac{d\sigma^{\lambda_e \lambda_p}_t}{dx dQ^2}}\; =\; \sum_q\,
{\frac{d\hat{\sigma}_t}{d\hat{t}}}^{\lambda_e \lambda_p}\, q^+(x,Q^2)\;
+\; {\frac{d\hat{\sigma_t}}{d\hat{t}}}^{\lambda_e -\lambda_p}\, q^-(x,Q^2)\; .
\eq

\no With $H2$ we get (${\hat \sigma}^{\lambda_e \lambda_p}_t\equiv 
{d\hat{\sigma}}^{\lambda_e \lambda_p}_t/{d\hat{t}}$) :
\EQ\label{eqAinter}
{\sigma}^{\lambda_e \lambda_p}_t\; \simeq\; {\hat \sigma}^{\lambda_e 
\lambda_p}_t u^+(x,Q^2)\, +\, {\hat \sigma}^{\lambda_e -\lambda_p}_t u^-(x,Q^2)\; .
\eq

\no For high values of $\Lambda$, the squared amplitudes for the CI 
($\sim 1/\Lambda^4$) are suppressed
compared to the interferences $\gamma\, CI$ ($\sim 1/Q^2\Lambda^2$) or
$Z\, CI$ ($\sim 1/(Q^2+M_Z^2)\Lambda^2$). In the following, for simplicity,
we will neglect the interference term $Z\, CI$ which is not dominant. \\
Then, we assume that the dominant terms for the CI come from
the interferences $\gamma\, CI$, given by eq.(\ref{TgaCI}).
In terms of the helicities, using $H1$, we deduce that the dominant
matrix elements squared are :
\EQA
T_{\gamma\, CI}^{--}\, (e^-)\; =\; \epsilon \,\frac{K}{2 \pi}\, {\hat s}^2\,
(1+\eta)(1+{\eta}')\; ,
\nonumber \\
T_{\gamma\, CI}^{++}\, (e^-)\; =\; \epsilon \,\frac{K}{2 \pi}\, {\hat s}^2\,
(1-\eta)(1-{\eta}')\; ,
\nonumber \\
T_{\gamma\, CI}^{--}\, (e^+)\; =\; \epsilon \,\frac{K}{2 \pi}\, {\hat s}^2\,
(1-\eta)(1+{\eta}')\; ,
\nonumber \\
T_{\gamma\, CI}^{++}\, (e^+)\; =\; \epsilon \,\frac{K}{2 \pi}\, {\hat s}^2\,
(1+\eta)(1-{\eta}') \; ,\label{Tchi}
\eqa
\no where $K={8\pi\alpha}/{3\Lambda^2 Q^2}$.
We find that the dominant cross sections for the subprocesses
are of the form ${\hat \sigma}^{\lambda \lambda}$
($\lambda =\lambda_e = \lambda_q$). We deduce that the dominant cross sections
for the process are :
\EQ\label{eqBi}
{\sigma_t}^{\lambda\lambda}\; \simeq\; {\hat \sigma}_t^{\lambda
\lambda} u^+ \;\;\;\;\;\;\;\;\;\;\;\;\; and \;\;\;\;\;\;\;\;\;\;\;\;\;\;
{\sigma_t}^{\lambda\, -\lambda}\; \simeq\; {\hat \sigma}_t^{\lambda
\lambda} u^- \; .
\eq
\no The hypothesis $H3$ indicates that the dominant process cross sections
are of the form ${\sigma_t}^{\lambda\lambda}$ (and not
${\sigma_t}^{\lambda -\lambda}$), and they correspond to :
\EQ
{\sigma_t}^{\lambda\lambda}\; \simeq\; {\hat \sigma}_t^{\lambda
\lambda} u^+ \; =\; \frac{\pi}{{\hat s}^2}\;
T_{\gamma\, CI}^{\lambda\lambda}\; u^+ \; .
\eq

\no We conclude that $\sigma_-^{--}$ is sensitive to $LL^\pm$,
$\sigma_-^{++}$ to $RR^\pm$, $\sigma_+^{--}$ to $RL^\pm$ and 
$\sigma_+^{++}$ to $LR^\pm$.\\
An immediate consequence is that the unpolarized cross sections
$\sigma_-$ and $\sigma_+$ are sensitive to ($LL^\pm$, $RR^\pm$) and 
($LR^\pm$, $RL^\pm$), respectively.

\no For the one-spin cross sections we get :
\EQ
{\sigma_t}^{\lambda}\; = \; \frac{1}{2}\, ({\sigma_t}^{\lambda\lambda} +
{\sigma_t}^{\lambda -\lambda})\; \simeq \; \frac{1}{2}\, {\hat \sigma}_t^{\lambda
\lambda} u
\eq
\no The sensitivity to each chirality is trivial. Moreover, the comparison of
this equation with eq.(\ref{eqBi}), indicates that the two-spin cross
sections are more sensitive to the CI than the one-spin ones, because :
\EQ
\frac{\sigma_t^{\lambda}}{\sigma_t^{\lambda\lambda}}\; \simeq\; \frac{u/2}{u^+} 
\; =\; 1 - \frac{\Delta u}{2 u^+}< 1
\;\;\;\;\;\;\;\;\;\;\; since \;\; \Delta u > 0\; .
\eq

The behavior of the different asymmetries is easily obtainable from
the above equations, because it is governed by its numerator,
since the denominators are dominated by the SM.
For instance, we obtain for the asymmetries considered in the text :
\EQ
num [\, A_{LL}^{PV} (e^-) \, ] \; \simeq \; \epsilon\, K\, (+\eta + \eta\prime )\;\; u^+\; ,
\eq
\EQ
num [\, A_{LL}^{PV} (e^+) \, ] \; \simeq \; \epsilon\, K\, (-\eta + \eta\prime )\;\; u^+\; ,
\eq
\EQ
num [\, B_2^2 \, ] \; \simeq \; -\, \epsilon\, K\, \eta (1- \eta\prime )\;\; u^+\; ,
\eq

\no and for the PC asymmetries (see eq.(\ref{eqBi})) :
\EQ
num [\, A_2^{1-4} \, ] \; = \; {\sigma_t}^{\lambda\lambda} - {\sigma_t}^{\lambda -\lambda}
\; \simeq\; {{\hat\sigma}_t}^{\lambda\lambda}\; \Delta u\; ,
\eq
\EQ
num [\, A_2^1 \, ] \; \simeq \; \epsilon\, \frac{K}{2}\, (1+\eta) (1+ \eta\prime )
\;\; \Delta u\; ,
\eq
\EQ
num [\, A_2^2 \, ] \; \simeq \; \epsilon\, \frac{K}{2}\, (1-\eta) (1- \eta\prime )
\;\; \Delta u\; ,
\eq
\EQ
num [\, A_2^3 \, ] \; \simeq \; \epsilon\, \frac{K}{2}\, (1-\eta) (1+ \eta\prime )
\;\; \Delta u\; ,
\eq
\EQ
num [\, A_2^4 \, ] \; \simeq \; \epsilon\, \frac{K}{2}\, (1+\eta) (1- \eta\prime )
\;\; \Delta u\; .
\eq
\no Then, $A_{LL}^{PV} (e^-)$ is sensitive to $(LL^\pm , RR^\pm )$, 
$A_{LL}^{PV} (e^+)$ to $(LR^\pm , RL^\pm )$, $B_2^2$ to $(RR^\pm , LR^\pm )$,
$A_2^1$ to $LL^\pm$, $A_2^2$ to $RR^\pm$,
$A_2^3$ to $RL^\pm$ and $A_2^4$ to $LR^\pm$.
\\

\vspace{2.cm}


\section{``Experimental asymmetries''/definition of 
$\Delta A_{stat}$}

\appendixB{}

\no Experimentally the particle beams are never fully polarized,
so we need to introduce some degree of polarization $P$.
Then, there is a shift between the values of the observable
defined theoretically ($P=1$) and the one measured experimentally ($P\ne 1$).\\
Only in the two cases $A_L$ and\footnote{$A^{PC}_{LL}$ is defined by
$A_{LL}^{PC} \equiv A_{LL}=({\sigma^{++} - \sigma^{+-}+\sigma^{--} -\sigma^{-+}})/(
{\sigma^{++} + \sigma^{+-} + \sigma^{--}+ \sigma^{-+}})$} 
$A_{LL}^{PC}$, the relation
between the two asymmetries ("theoretical" and "experimental")
is simple, and a suitable redefinition
of the measured asymmetry makes the situation clear (see below). In some
other cases, like the double PC spin asymmetry $A_\parallel$ measured in
polarized deep inelastic scattering {\it at low $Q^2$}, some
relevant assumptions (parity conservation)
allow to obtain a similar simple relation (see below).
However, in general, we do not have such simple relations between
the two asymmetries, so we need a careful treatment,
in order to take into account the degrees of polarization,
and to define the statistical error of the asymmetries correctly.\\

The goal of this appendix is to show the procedure that we have
used in the present article. This can be used in polarized
hadronic collisions as well.
Since it is relatively technical, we
introduce our notations, then, we illustrate our general procedure on
$A_{LL}^{PV}$, since it is valid for any spin asymmetries. Finally,
we give two examples which reproduce the simple relations mentioned 
above.

\subsection{Notations}

\no We note $\sigma^{\lambda_a \lambda_b}\; (\equiv 
\frac{d\sigma^{\lambda_a \lambda_b}}{dXdY...})$ the cross section
corresponding to the process  
${\vec a}(\lambda_a)\; {\vec b}(\lambda_b)\, \rightarrow\, X$.\\
The basic observables are : $\sigma^{++},\; \sigma^{+-},\;
\sigma^{-+}$ and $\sigma^{--}$.
They are related to the one-spin cross sections by 
$\sigma^{\lambda_a\, 0}\; =\; \frac{1}{2}\; 
(\, \sigma^{\lambda_a\: \lambda_b}\, +\,
\sigma^{\lambda_a\, - \lambda_b} \, )$, when
only $\vec{a}$ are polarized, and similarly 
$\sigma^{0\lambda_b} = \frac{1}{2} 
( \sigma^{\lambda_a \lambda_b} +
\sigma^{-\lambda_a \lambda_b} )$, when
only $\vec{b}$ are polarized.
The unpolarized cross section is :
\EQ
\sigma^{0 0} = \frac{1}{4} ( \sigma^{--} + \sigma^{++} 
+ \sigma^{-+} + \sigma^{+-} )
\; =\; \frac{1}{2} ( \sigma^{-0} + \sigma^{+0}) 
\; =\; \frac{1}{2} ( \sigma^{0-} + \sigma^{0+})\; .
\eq

\no Concerning the numbers of events, for a particular spin
configuration ($\lambda_a ,\lambda_b$), we define
$N^{\lambda_a\lambda_b}\; =\; L^{\lambda_a\lambda_b}\, \times\, 
\sigma^{\lambda_a\lambda_b}$. We make the special choice : 
$L^{\lambda_a\lambda_b}\, =\, \frac{1}{2}\, L^{\lambda_a 0}\,
=\, \frac{1}{2}\, L^{0\lambda_b}\,=\, \frac{1}{4}\, L^{0 0}$,
where $L^{00}$ is the ``total'' integrated luminosity.
Then we have :
\EQ
N^{\lambda_a\, 0}\; =\; N^{\lambda_a\: \lambda_b}\, +\, N^{\lambda_a\, - \lambda_b} 
\;\;\;\;\;\; , \;\;\;\;\;\;
N^{0\, \lambda_b}\; =\; N^{\lambda_a\: \lambda_b}\, +\, 
N^{- \lambda_a\, \lambda_b} \; ,
\eq
\EQ
N^{0 0}\; =\; N^{--}\, +\, N^{++}\, +\, N^{-+}\, +\, N^{+-} \;
=\; N^{-0}\, +\, N^{+0} \; =\; N^{0-}\, +\, N^{0+}\; .
\eq
\\

\no Now, if we define an asymmetry by :
\EQ
A\; =\; \frac{N_1 \, -\, N_2}{N_1 \, +\, N_2}\;\;\; ,
\eq
\no its statistical error is \cite{PDG} :
\EQ
\Delta A
\; =\; \sqrt{\frac{1-A^2}{N_1+N_2}}
\; =\; \frac{2}{(N_1 + N_2)^2}\, \sqrt{N_1 N_2 (N_1 + N_2)}
\; =\; \frac{1-A^2}{\sqrt{1-A}}\frac{1}{\sqrt{2\, N_1}}
\eq

\subsection{General procedure : example on \ALLPV}

Consider the process ${\vec a}\; {\vec b}\, \longrightarrow\, X$.
Experimentally, we have a fraction $P_a$ of particles $a$ with helicity
$\lambda_a$, and a fraction $1-P_a$ of unpolarized $a$ particles, which
collide with a fraction $P_b$ of particles $b$ with helicity
$\lambda_b$, and a fraction $1-P_b$ of unpolarized $b$ particles.
We obtain the "experimental" cross section :
\EQA
\hspace*{-1.2cm}
\sigma_{exp}^{\lambda_a \lambda_b}\; &=& \; P_a\,P_b\, \sigma^{\lambda_a \lambda_b}\, 
+\, P_a\, (1-P_b)\, \sigma^{\lambda_a 0}\, + (1-P_a)\, P_b\, 
\sigma^{0 \lambda_b}\, +\, (1-P_a)\, (1-P_b)\, \sigma^{00} \\
%
&=& \; 
\frac{1}{4}\, (1+P_a)\, (1+P_b)\, \sigma^{\lambda_a \lambda_b} \, +\, 
\frac{1}{4}\, (1-P_a)\, (1-P_b)\, \sigma^{-\lambda_a -\lambda_b} \nonumber \\
&&\label{def2exp} +\,
\frac{1}{4}\, (1+P_a)\, (1-P_b)\, \sigma^{\lambda_a -\lambda_b} \, +\, 
\frac{1}{4}\, (1-P_a)\, (1+P_b)\, \sigma^{-\lambda_a \lambda_b} 
\eqa

\no In terms of events, replace in the preceding equation (B.7), $\;\sigma$
by $N$.\\
\no We mean by ``experimental'' that the observable (${\cal O}_{exp}$
with ${\cal O} \equiv \sigma , \, A, \, B\, ...$) is directly dependent
on the degrees of polarization. In fact, ${\cal O}_{exp}$ could
correspond, on the one hand, to a theoretical
quantity which takes into account the expected experimental
conditions, or on the other hand, to a measured quantity in some actual
experimental conditions (${\cal O}_{exp}\equiv {\cal O}_{meas}$). 
The former case is involved in any phenomenological
simulation which intend to take into account the degrees of polarization,
and we have to use eq.(B.7) to define the statistical error of the
theoretical asymmetry properly. In the latter case, in order to reconstruct
the number of events which are independent on the degrees of polarization,
from the measured number of events, we will have to use the inverse formula
of eq.(B.7) :
\EQA\hspace*{-0.4cm}
\sigma^{\lambda_a \lambda_b} &=&  
\frac{1}{P_a^{meas} P_b^{meas}} \left[
\frac{1}{4} (1+P_a^{meas}) (1+P_b^{meas}) 
 \sigma_{meas}^{\lambda_a \lambda_b}  + 
\frac{1}{4} (1-P_a^{meas}) (1-P_b^{meas}) 
\sigma_{meas}^{-\lambda_a -\lambda_b} \right.
\nonumber \\
&&\hspace*{-1.cm}-\,\left.
\frac{1}{4}\, (1+P_a^{meas})\, (1-P_b^{meas})\, 
\sigma_{meas}^{\lambda_a -\lambda_b} \, -\, 
\frac{1}{4}\, (1-P_a^{meas})\, (1+P_b^{meas})\, 
\sigma_{meas}^{-\lambda_a \lambda_b} \right]
\eqa
\no We insist on the fact that, now, $\sigma^{\lambda_a \lambda_b}$
is not a cross section given by theory, but a cross section constructed
from the measured cross sections to be $P_a,P_b$ independent.\\


\no $\bullet$ \underline{Example on \ALLPV}\\

We begin this example in the framework of phenomenological simulations.
In this case, $P_a$ and $P_b$ are the {\it expected} degrees of polarization
for the beams $a$ and $b$. \ALLPV and $N^{--}$, $N^{++}$, $N^{-+}$ and $N^{+-}$ 
are some quantities predicted by theory, so, they are
$P_a$, $P_b$ {\it independent}. We call
$A_{LL}^{PV}{\scriptstyle (exp)}$ and
$N^{--}_{exp}$, $N^{++}_{exp}$, $N^{-+}_{exp}$ and $N^{+-}_{exp}$ 
some {\it theoretical} quantities which are $P_a$, $P_b$ {\it dependent}.

We call ``theory 1'' the strategy which consists to
use as the basic observable $A_{LL}^{PV}{\scriptstyle (exp)}$. In this case, 
with eq.(B.7), we have :

\EQ\hspace*{-10.2cm}
A_{LL}^{PV}{\scriptstyle (exp)}
=\; \frac{N_{exp}^{--} \, -\, N_{exp}^{++}}
{N_{exp}^{--} \, +\, N_{exp}^{++}}
\eq
\EQ\label{apvexperso}
= \frac{
\frac{1}{2} (P_a + P_b) \left[N^{--} - N^{++} \right]
 + \frac{1}{2} (P_a - P_b) \left[N^{-+}- N^{+-} \right]}
{\frac{1}{2} \left[N^{--} + N^{++} + N^{-+} 
+ N^{+-} \right]
+ \frac{1}{2} P_aP_b \left[N^{--} +N^{++} - N^{-+}  
- N^{+-} \right]}
\eq

\no From this equation, we see that $A_{LL}^{PV}{\scriptstyle (exp)}$ is not 
simply related to $A_{LL}^{PV}$. It means that, even if 
$A_{LL}^{PV}{\scriptstyle (exp)}$ is defined with the ("experimental") number
of events for \underline{two} spin configurations, the presence of the degrees
of polarization induces an admixture of the (theoretical) number
of events for the \underline{four} spin configurations.\\
However, this asymmetry is well-defined and its statistical error too :

\EQ
\Delta A_{LL}^{PV}{\scriptstyle (exp)} = 
\sqrt{\frac{1-(A_{LL}^{PV}{\scriptstyle (exp)})^2}
{N_{exp}^{--} \, +\, N_{exp}^{++}}} \; .
\eq
\\

\no Now, we call ``theory 2'' the strategy which consists to
use as the basic observable $A_{LL}^{PV}$, which is simply defined by :
\EQ
A_{LL}^{PV} = \frac{N^{--} - N^{++}}{N^{--} + N^{++}} \; .
\eq

\no Actually, as in ref.\cite{Prescott} for the study of $A_L$, 
we want \ALLPV to be the basic observable of
the analysis.
So the question arise : what is its statistical error ?
The answer boils into the fact that we want the two methods, 
theory 1 and theory 2,
to be equivalent, or practically, to have the same analyzing power
({\it i.e.} the same $\chi^2 \sim (A/\Delta A)^2$). Namely, we require that :
\vspace{2.mm}
\EQ
\frac{A}{\Delta A}\, =\, \frac{A^{exp}}{\Delta A^{exp}} \; .
\eq

\no So, we define the statistical error of the theoretical asymmetry
in the following way :
\EQ
\Delta A_{LL}^{PV}\, =\, \frac{A_{LL}^{PV}}
{A_{LL}^{PV}{\scriptstyle (exp)}} .
\Delta A_{LL}^{PV}{\scriptstyle (exp)} \; .
\eq
\\

Now, if we are in the framework of some experimental measurements, we have
at hand some {\it measured} quantities :
$P_a^{meas}$, $P_b^{meas}$,
$N^{--}_{meas}$, $N^{++}_{meas}$, $N^{-+}_{meas}$ and $N^{+-}_{meas}$.
From these ones, we can construct (using eq.(B.8)) the 
$P_a,P_b$ independent number of events 
$N^{--}$, $N^{++}$, $N^{-+}$ and $N^{+-}$.

Then, we call ``measure 1'' the strategy which consists to
define as the basic observable an asymmetry which is {\it dependent} on the
degrees of polarization. In this case, we have simply :
\EQ
\left(
A_{LL}^{PV}{\scriptstyle (exp)}\right)_{meas} 
= \frac{N^{--}_{meas} - N^{++}_{meas}}
{N^{--}_{meas} + N^{++}_{meas}}
\eq
\EQ
 \left(
\Delta A_{LL}^{PV}{\scriptstyle (exp)}\right)_{meas} = 
\sqrt{\frac{1-(A_{LL}^{PV}{\scriptstyle (exp)})_{meas}^2}
{N_{meas}^{--} \, +\, N_{meas}^{++}}}
\eq

\no We have to compare this asymmetry to the one defined in ``theory 1''.\\

\no We call ``measure 2'' the strategy which consists to
define as the basic observable an asymmetry which is {\it independent} on the
degrees of polarization. Then, we construct \ALLPV by the use of
eq.(B.8), and we find :
\EQA\hspace*{-0.5cm}
A_{LL}^{PV} &=& \frac{
\frac{1}{2P_a^{meas}}\left[N^{--}_{meas} + N^{-+}_{meas}-N^{+-}_{meas} - 
N^{++}_{meas} \right]
 + \frac{1}{2P_b^{meas}}\left[N^{--}_{meas} - N^{-+}_{meas}
+N^{+-}_{meas} - N^{++}_{meas} \right] }
{\frac{1}{2}\left[N^{--}_{meas} + N^{-+}_{meas}
+N^{+-}_{meas} + N^{++}_{meas} \right]
 + \frac{1}{2P_a^{meas}P_b^{meas}}\left[N^{--}_{meas} - N^{-+}_{meas}
-N^{+-}_{meas} + N^{++}_{meas} \right] }
\nonumber \\
&=& \; \frac{
\frac{1}{2P_a^{meas}} \left[N^{-0}_{meas} \, -\, N^{+0}_{meas} \right]
\, +\, \frac{1}{2P_b^{meas}} \left[N^{0-}_{meas} \, -\, N^{0+}_{meas} \right]}
{\frac{1}{2}\, N^{00}_{meas} \, (\, 1 \, +\, \frac{1}
{P_a^{meas}\,P_b^{meas}}\,  A_{LL}^{meas}\, )}
\eqa
\no The associated statistical error is :
\EQ
\Delta A_{LL}^{PV}
\, =\, \frac{A_{LL}^{PV}}
{(A_{LL}^{PV}{\scriptstyle (exp)})_{meas}} .
(\Delta A_{LL}^{PV}{\scriptstyle (exp)})_{meas}
\eq

\no We have to compare this asymmetry to the one defined in ``theory 2''.\\

\no The two procedures, ``experimental'' ({\it i.e.}
theory/measure 1 : $P$ dependent) and ``theoretical'' ({\it i.e.}
theory/measure 2 : $P$ independent), are
physically equivalent. This remark is valid for phenomenological
simulations and for experimental measurements. We can choose
the presentation of the results that we prefer.
In order to present some curves which are independent
of the assumed degrees of polarization, we have to use the strategy that
we have called ``theoretical''.
We have followed this procedure in this paper.\\
The strategies presented for \ALLPV are valid for any spin asymmetries.\\

\no Remark on the \ALLPV case :\\
\no From the equation (\ref{apvexperso}), we have seen that there is no 
simple relation between
$A_{LL}^{PV}{\scriptstyle (exp)}$ and $A_{LL}^{PV}$.
However, if we make the crude assumption that $A_{LL}=0$ 
({\it i.e.} $P_aP_bA_{LL}=0\, and\, \frac{1}{2}\, N^{00}
=N^{--} + N^{++} = N^{-+} + N^{+-}$), and if we assume $P=P_a=P_b$, then 
we obtain the simple relation :
\EQ\label{dasapv}
A_{LL}^{PV}(exp)\; = \;P\, \; A_{LL}^{PV}\;\;\;
\longrightarrow \;\; \Delta A_{LL}^{PV}\; =\; \frac{1}{P}\;
\sqrt{\frac{1-P^2\, A_{LL}^{PV\, 2}}{N^{--} \, +\, N^{++}}}
\eq

\no We have suppressed the label {\small\it ``meas''},
because when there is a ``simple relation'' between the two asymmetries,
the distinction between phenomenological
simulations and experimental measurements,
is no more required since they obey
the same equation (here eq.(\ref{dasapv})), even if physically the
distinction between the two conditions still exists.\\
Finally, note that eq.(\ref{dasapv}) is obtained under a crude assumption,
but it exhibits the interesting property that the statistical errors of the
PV spin asymmetries are $\sim 1/P$. This has been already noted
in \cite{ptjmvctprd}, and in \cite{kamal}, where the formula
eq.(B.17) has been also derived.\\


\subsection{Connections with some usual spin asymmetries}

The procedure described above is valid for any spin asymmetries.
However, if for a certain asymmetry (say $A$), there is a "simple
relation" between the theoretical definition ({\it i.e.} $A$, $P$
independent) and the experimental definition ({\it i.e.} $A_{exp}$, $P$
dependent), it appears that the statistical error of the theoretical
asymmetry has a simple expression. We examplify this fact for the
two usual asymmetries $A_{LL}$ and $A_\parallel$.\\
\newpage

\no $\bullet$ {\bf {\ALL }}\\

\no The $P$ dependent experimental asymmetry is defined by :
\EQ
A_{LL}^{exp}\; =\; \frac{N_{exp}^{++} \, -\, N_{exp}^{+-}
\, -\, N_{exp}^{-+} \, +\, N_{exp}^{--}}
{N_{exp}^{++} \, +\, N_{exp}^{+-}
\, +\, N_{exp}^{-+} \, +\, N_{exp}^{--}}\;\;\; ,
\eq

\no associated to the statistical error :
\vspace{2.mm}
\EQ
\Delta A_{LL}^{exp}\; =\; \sqrt{\frac{1-A_{LL}^{exp\, 2}}
{N_{exp}^{++} \, +\, N_{exp}^{+-}
\, +\, N_{exp}^{-+} \, +\, N_{exp}^{--}}}\;\;\; .
\eq

\no Now, using the formula of correspondence (eq.(B.7)), 
we obtain the well-known simple formula : $A_{LL}^{exp}\; =\; P_a\,P_b\; A_{LL}$.
\\

\no Then, we deduce the statistical error for the
theoretical $A_{LL}$ :
\EQ
\Delta A_{LL}\, =\, 
\frac{1}{P_a\, P_b} 
.\Delta A_{LL}^{exp}
 =\; \frac{1}{P_aP_b}\; \sqrt{\frac{1-A_{LL}^{exp\, 2}}
{N_{exp}^{00}}}
\; =\; \frac{1}{P_aP_b}\; \sqrt{\frac{1-P_a^{\, 2}P_b^{\, 2}\, A_{LL}^{\, 2}}
{N^{00}}}
\eq

\no It is clear on this example, that it is no more necessary to 
distinguish the fact that we are performing a phenomenological analysis
or analysing some experimental measurements. The equations are the same
for the two cases (for the latter case and using our notations,
 we just have to add the subscript
${\scriptstyle meas}$ to all the preceding equations).
\\

\no $\bullet$ {\bf {$A_\parallel$ }}\\

\no In general, in polarized deep inelastic scattering, rather
than analysing the spin asymmetry \ALL it is usual
to study its reduced expression $A_\parallel (\equiv A_2^2$ or $A_2^4)$.\\
\no The $P$ dependent experimental asymmetry is defined by :
\EQ
A_\parallel {\scriptstyle (exp)}\; =\; \frac{N_{exp}^{++} \, -\, N_{exp}^{+-}}
{N_{exp}^{++} \, +\, N_{exp}^{+-}}\;\;\; ,
\eq

\no associated to the statistical error :
\vspace{2.mm}
\EQ
\Delta A_\parallel {\scriptstyle (exp)}\; =\; \sqrt{\frac{1-
 A^2_\parallel 
{\scriptstyle (exp)}}
{N_{exp}^{++} \, +\, N_{exp}^{+-}}}\;\;\; .
\eq

\no In terms of the ``theoretical'' number of events, we obtain :
\vspace{2.mm}
\EQ
A_\parallel {\scriptstyle (exp)}\; = \; \frac{
\frac{1}{2}\, (1+P_a)\, P_b\, \left[N^{++} \, -\, N^{+-} \right]
\, -\, \frac{1}{2}\, (1-P_a)\, P_b\, \left[N^{--} \, -\, N^{-+} \right]}
{\frac{1}{2}\, (1+P_a)\, \left[N^{++} \, +\, N^{+-} \right]
\, +\, \frac{1}{2}\, (1-P_a)\, \left[N^{--} \, +\, N^{-+} \right]}
\eq

\no We see, now, that there is no more a simple relation
between $A_\parallel {\scriptstyle (exp)}$ and $A_\parallel $.
However, if we assume parity conservation
({\it i.e.} $N^{++} = N^{--}$ and 
$N^{+-} = N^{-+}$), we recover
the well-known formula : $A_\parallel {\scriptstyle (exp)}
\; =\; P_a\,P_b\; A_\parallel $,
giving for the ``theoretical'' statistical error :
\EQ\label{aparr}
\Delta A_\parallel \, \simeq \, 
\frac{1}{P_a\, P_b} 
. \Delta A_\parallel {\scriptstyle (exp)}\;
=\; \frac{1}{P_a\, P_b}\; \sqrt{\frac{1-P_a^{\, 2}P_b^{\, 2}\, 
A^2_\parallel }{N^{++} \, +\, N^{+-}}}
\eq

\no If we do not assume parity conservation, 
eq.(\ref{aparr}) is no longer valid, and we have to use 
the procedure presented for $A_{LL}^{PV}$.\\
Finally, we have used a similar procedure in the one-spin case.\\


\end{document}